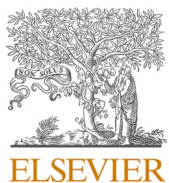

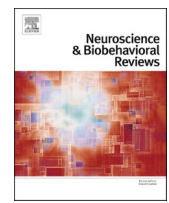

# How higher goals are constructed and collapse under stress: A hierarchical Bayesian control systems perspective

Rutger Goekoop [a],[*], Roy de Kleijn [b]

[a] Parnassia Group, PsyQ, Department of Anxiety Disorders, Early Detection and Intervention Team (EDIT), Netherlands
[b] Cognitive Psychology Unit, Leiden University, Netherlands

## ARTICLE INFO

## ABSTRACT

In this paper, we show that organisms can be modeled as hierarchical Bayesian control systems with *small world* and information bottleneck (*bow-tie*) network structure. Such systems combine hierarchical perception with hierarchical goal setting and hierarchical action control. We argue that hierarchical Bayesian control systems produce deep hierarchies of goal states, from which it follows that organisms must have some form of 'highest goals'. For all organisms, these involve internal (self) models, external (social) models and overarching (normative) models. We show that goal hierarchies tend to decompose in a top-down manner under severe and prolonged levels of stress. This produces behavior that favors short-term and self-referential goals over long term, social and/or normative goals. The collapse of goal hierarchies is universally accompanied by an increase in entropy (disorder) in control systems that can serve as an early warning sign for tipping points (disease or death of the organism). In humans, learning goal hierarchies corresponds to personality development (maturation). The failure of goal hierarchies to mature properly corresponds to personality deficits. A top-down collapse of such hierarchies under stress is identified as a common factor in all forms of episodic mental disorders (psychopathology). The paper concludes by discussing ways of testing these hypotheses empirically.

## 1. Introduction

For centuries, scientists have attempted to discover natural laws that govern the structure and function of living systems. This effort is now producing some interesting results due to theoretical advances, the advent of high-throughput datasets and a huge increase in computing power (Kitano, 2017). Currently, the field still shows a global division between biological and computer sciences, which represents a fundamental distinction in the way the problem has been approached to date, i.e. either by studying living systems themselves (e.g. biology, genetics, biochemistry) or by studying artificial versions of them (e.g. engineering, computer science and robotics). Below, we will first discuss progress in the fields of artificial systems and biological systems separately. We will then merge insights from both fields to produce a general theory on information processing in living systems and the way they respond to stress. We highlight the universality of this response along with its applicability in humans, and conclude by discussing methods to test the model empirically.

## 2. Artificial systems

### 2.1. Organisms as control systems

Artificial intelligence has now come to a point where computers are able to reach (super)human level performance in complex tasks without prior instructions (Mnih et al., 2015; Schmidhuber, 2015; Silver et al., 2017). The basis for this achievement lies in the beginning of the 20th century, when cyberneticists such as W.E. Ashby began to model organisms as control systems (Ashby, 1961; Cannon, 1929, 1932; Powers, 1973a). Such systems maintain internal stability despite changes in environmental conditions by generating some kind of output (O) that aims to match the current input state (I) with a desired or anticipated throughput state (T; a reference value or setpoint). Ashby was the first to highlight the importance of a close coupling between the output and the input of such systems, which is referred to as 'feedback'. To prove his claims, Ashby constructed a device called a 'homeostat', which involved four subsystems that kept each other in check. Each subsystem consisted of a first-order feedback loop that regulated 'essential variables' (e.g. blood pressure, glucose levels) and a second-order feedback loop that re-organized a system's input–output relations when first-order feedback failed, allowing the perturbed system to revert to a stability of its essential variables after all (Seth, 2014). The former, primary form of stability is referred to as homeostasis, whereas the latter form of stability (through additional change) is referred to as 'allostasis'. This double feedback system is one of the earliest forerunners of 'hierarchical control' (see below). By combining four coupled subsystems into one homeostat, the entire control system showed 'ultrastable' behavior.

---

* Corresponding author at: Parnassia Group, PsyQ, Department of Anxiety Disorders, Early Detection and Intervention Team (EDIT), Lijnbaan 4, 2512VA The Hague, Netherlands.
*E-mail address:* R.Goekoop@psyq.nl (R. Goekoop).

In engineering, control systems are used e.g. in central heating systems, which aim to maintain a stable room temperature despite environmental fluctuations by controlling the radiator. This is done using a control system that compares the current room temperature encoded by a temperature sensor (an input node) to that of a thermostat, which serves as a reference node that encodes a desired temperature (a setpoint). The difference between the two (the error) is transferred in some form to the radiator (an output node), which tries to close the gap between the desired and actual room temperatures (the environment) by emitting heat. Studies indicate that living systems have conditionally independent compartments for input, evaluation and output that allow them to behave in similar ways as control systems (Kirchhoff et al., 2018). Organisms use their senses to monitor the state of their environment and compare their input states to a setpoint state located within a throughput part. The error is then transferred to the output part of the organism, which tries to close the gap between the desired and actual environmental states by generating action (see Fig. 1). Actions change the state of environment, which feeds back into the senses and the process is repeated. This iterative process helps organisms to find an optimal environmental niche. For example, motor activity in woodlice continues almost ceaselessly and drops to zero only when humidity levels reach near 100 % (a setpoint). As a result, woodlice keep running around erratically until they hit upon a wet place, which is why we find these creatures in all sorts of nooks and crannies. This behavior helps woodlice prevent desiccation and makes them invisible to predators (Friston et al., 2018).

Seminal work by W.T. Powers (1973a,b) showed that biological systems vary their output freely until the state of the input node matches a reference value. Their behavior thus serves to keep a percept (of some environmental condition) within certain limits. Woodlice probably have no clue as to where exactly in the garden they can find a particular crevasse, after which they engage in a carefully controlled output sequence that is aimed at reaching the desired spot. Instead, they just stumble upon a dark and wet place that produces the kind of sensor output that makes motor activity drop to zero. Since Powers considered organisms to control their input (percepts) by means of their output (behavior) and its subsequent effects on the environment, this type of control was called 'perceptual control' (Powers, 1973a). Perceptual control theory is highly pragmatic: rather than the specific actions, it's the end-result that counts. By freely 'emitting behavior' (Skinner, 1990) until a desired effect is obtained, organisms can come up with a number of different solutions to the same problem (e.g. running and hiding in crevasses, rolling up, or digging in all prevent desiccation). This adds flexibility and creativity to the production of behavior (Powers, 1973b). The advantages of perceptual control have been demonstrated in a number of experiments. For instance, robots that run on perceptual control systems can be pushed off their feet in many different ways yet remain stable, whereas robots that run solely on action-control systems can correct their position only in a limited number of ways and tip over (Johnson et al., 2020).

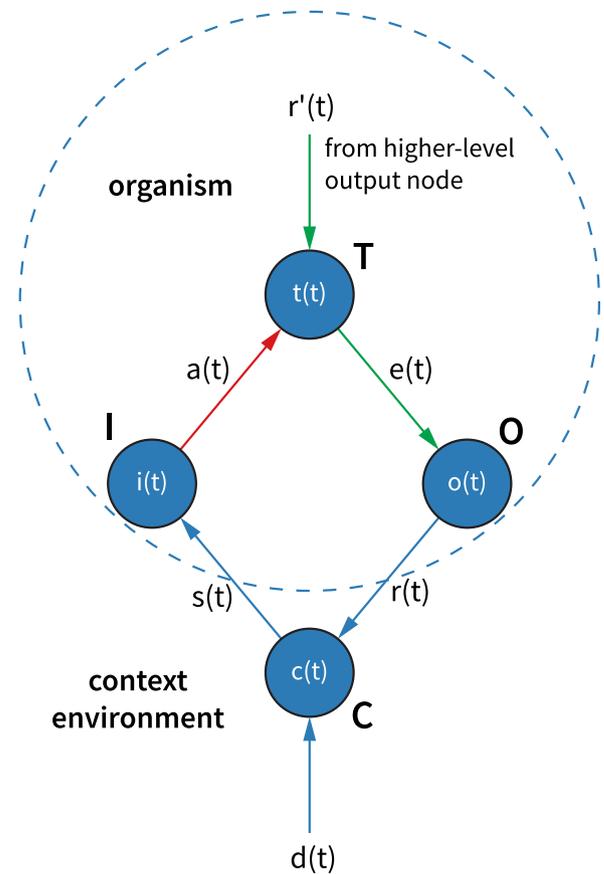

**Fig. 1.** Organisms as Control Systems.
*Note:* Organisms can be modeled as control systems that consist of an input node I (a sensor), a throughput node T (a setpoint) and an output node O (an effector), which are connected by links that symbolize the possibility of energy exchange between these nodes (see text). Arrows show the direction of energy flow, colors indicate positive or negative relationships (red: negative, blue: positive). The sensory node I has a state $i(t)$ that is changed as a result of a stimulus $s(t)$ from the environment C (context), which is in a changing state c (t). The state $i(t)$ of the sensory node I is sampled by an afferent connection and the resulting state $a(t)$ is compared to (i.e. subtracted from) the state T(t) of a throughput node T (the setpoint or reference node). The difference (error $e(t)$) between the two states is passed on by efferent connections to the output node O (in state $o(t)$), which generates the corrective response $r(t)$ to the environment C, and so on. External disturbances of the environment C are modeled by $d(t)$. The setpoint of the system T can be reset by the output from higher level control systems, see text.

### 2.2. Organisms as hierarchical control systems

Graphical models such as Fig. 1 can produce behavior that can appear quite life-like (Braitenberg, 1984; Powers, 1973a). Nevertheless, such models require an extension in order to explain more complex forms of behavior, i.e. the formation of action sequences that allow organisms to accomplish more complex tasks. For instance, making a cup of coffee involves a number of simple subtasks ('action primitives') that need to be placed in a particular order in order to succeed (e.g. heating water, grabbing a cup, pouring hot water over churned coffee beans, pouring the coffee into a cup, adding milk or sugar, etcetera; Botvinick, 2007). Such output sequences can be more or less efficient depending on the order and the number of recursions in which the subtasks appear (Commons and Pekker, 2008; Solway et al., 2014). Powers showed that perceptual control systems can produce action sequences (behavior) by allowing their setpoints to be reset by the output of other (higher level) control systems and so on, yielding a hierarchy of control systems (Powers, 1973a,b; Powers et al., 1960). This hierarchy is symbolized in Fig. 1 by the input r'(t) to the setpoint node T. This is the output of a higher-level control system that feeds into the reference signal of a subordinate control system. Fig. 2 shows a more elaborate example of a hierarchical perceptual control system.

The idea that higher order control systems continuously update the setpoints of lower-level systems (to eventually affect the activity of action primitives) is known as the *equilibrium setpoint hypothesis* for motor control (Feldman and Levin, 2009). In hierarchical action control, primitives can be compared to individual musical notes that are activated in parallel ('chords') or in different sequences in order to produce the 'music' of behavior. Studies of hierarchical action control show that action sequences do not require a strict correspondence with the hierarchical wiring of the control system (i.e. we do not engage in a fully hierarchically controlled sequence of coffee-making actions that is spat out from the very beginning of grabbing coffee beans to sipping from the cup; de Kleijn et al., 2014d). Rather, organisms produce intermittent bursts of hierarchically organized action sequences that are updated by a repeated sampling of the environment (action–perception cycles) (Botvinick, 2007). This is comparable with a musician looking up and down at the scroll sheet every now and then to keep track of the piece. Although hierarchical *action* control seems to contradict the notion of hierarchical *perceptual* control, it remains consistent with this notion in the sense that organisms use their (hierarchically controlled) action sequences to eventually control their input states via the environment. Hierarchical action control is routinely used in e.g. robotics, allowing robotic systems to show complex forms of behavior (e.g. Brooks, 1986).

In the past few decades, graphical models of control systems have been modified to explain increasingly complex forms of behavior. Much progress came from studies of reinforcement learning (also termed *operant conditioning*), which added the elements of memory and prediction to control systems (Jordan and Mitchell, 2015; Sutton and Barto, 2018). Such systems update their policies (input–output strategies) depending on the expected reward of some action. The expected reward (a prediction) is encoded by the setpoints of these systems, of which the state represents the reward or value obtained after a previous action (i.e. a memory). These predictive setpoints are continuously reset (updated) as a function of previous outcomes, keeping track of the values that maximized reward in the past. Thus, reinforcement learning systems iteratively learn the policies that maximize long-term cumulative reward. Whereas earlier systems made no detailed models of the environments they live in (so called *model-free* systems), later systems were allowed to make explicit predictions of the way in which certain imaginary actions would change the input to the system, considering previous experiences (*model-based* systems) (Doll et al., 2012; Solway and Botvinick, 2012). Such 'world models' are simulations of actions and their possible outcomes (e.g. where different paths in a maze lead to and how rewarding that would be), which are based on memories of previous actions and their outcomes. Predictive activity of this type has been compared to the act of planning, imagination, or goal setting, which is why model-based systems are alternatively referred to as *goal-directed* systems. Goal-directed systems require an elaboration of their throughput parts, to accommodate hierarchies of setpoints that encode complex predictive models of the world. Such 'goal states' are continuously updated and pursued by hierarchically organized action sequences until a maximum value has been reached. Studies show that hierarchical model-based systems such as these outperform hierarchical model-free systems in spatial navigation tasks (Botvinick and Weinstein, 2014). This is because such systems construct hierarchies of goals and corresponding subgoals (so called 'goal hierarchies'), which are each pursued in a logical order until the global goal has been reached (e.g. 'get to fruit' = climb tree, jump to other tree, sling to branch, grab the fruit, eat the fruit).

In the past decade, goal-directed learning has been applied within the context of artificial neural networks (Schmidhuber, 2015). Such networks consist of a layer of input nodes that connect to a layer of throughput nodes (a *hidden* layer), which in turn connects to a layer of output nodes. When such systems are trained, the connections within the network are altered until a given input produces a suitable output. It turns out that the performance of such systems increases significantly when their throughput parts are extended to include multiple, hierarchically ordered layers of nodes. Such deep networks can associate raw perceptual input (say, the image of a cat) to a suitable output (e.g. a hierarchical output sequence 'C – A – T') with remarkable precision. When deep networks are allowed to construct explicit world models

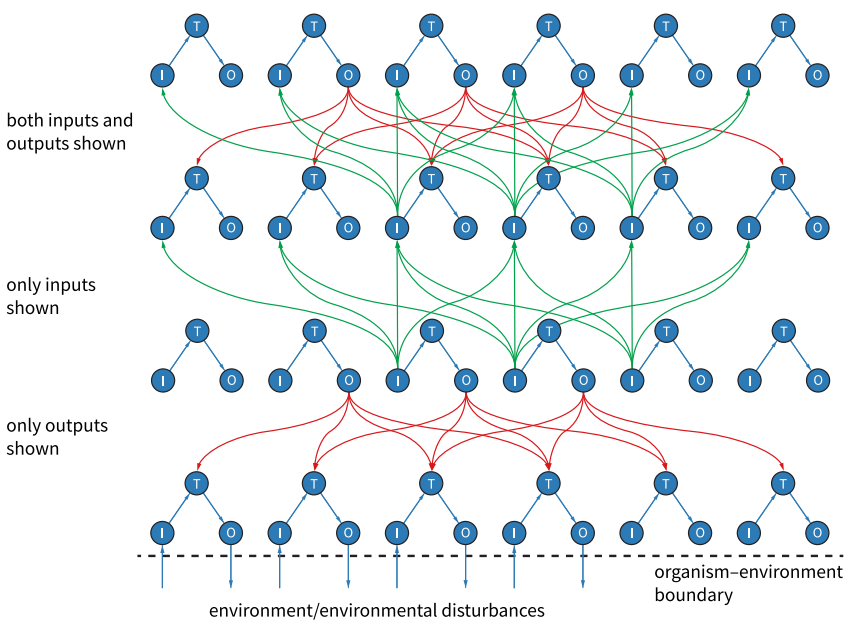

**Fig. 2.** An Example of a Hierarchical Perceptual Control System.
*Note:* Classical example of an artificial hierarchical control system, which involves the stacking of one control system on top of another, to produce multiple levels of control. This can be compared to the stacking of one array of thermostats on top of another in order to better control temperature fluctuations in the environment. The output of higher-level control systems can modify the setpoints of subordinate systems (and so on) to produce ordered sequences of action primitives, which we call behavior.

both inputs and outputs shown

only inputs shown

only outputs shown

environment/environmental disturbances

organism–environment boundary

(goal hierarchies), their performance increases even further. Such hierarchical 'deep belief' systems can 'imagine' a future and formulate efficient sequences of goals and corresponding subgoals that are pursued by means of complex action sequences until the input to the system matches the global goal. Such systems can achieve high success rates (Nagabandi et al., 2018; Pascanu et al., 2017; Racanière et al., 2017; Yamins and DiCarlo, 2016). The performance of these systems comes close to what neuroscientists believe is the essential nature of the human brain: an active inference engine, whose primary job it is to construct predictive models of what is going on in the environment and to test these models by performing some kind of action out into the environment. Such actions change the input to the system (via the environment), which serves as a check on model evidence (Friston, 2010. According to active inference theory, organisms cannot only reduce prediction errors by varying motor output impacting on percepts (as perceptual control would have it), but also by updating their world models to produce a better fit with their input states (a process called 'Bayesian belief updating'). See below for further information on active inference.

In summary, adding hierarchy to the output parts of control systems allows for the production of complex action-perception sequences (behavioral hierarchies), whereas adding hierarchy to the throughput (goal) parts further boosts the performance of such systems by producing efficient strategies (goal hierarchies). More recently, studies began to apply hierarchical structure to the input layers of deep networks (Mnih et al., 2015; Simonyan and Zisserman, 2014). The hierarchical structure of perceptive areas has been relatively ignored in previous studies, despite the fact that this is a well-known attribute of the cerebral cortex in higher mammals (e.g. receptive fields in the macaque visual cortex) (Hegdé and Felleman, 2007; Rohe and Noppeney, 2015). Hierarchical perception allows control systems to extract increasingly abstract patterns and shapes from raw perceptual input (Karklin and Lewicki, 2009; Kriegeskorte, 2015; Tenenbaum et al., 2011). In 2015, a seminal study was the first to combine hierarchical input (abstract vision) with hierarchical throughput (abstract goal-setting) and hierarchical output (complex action, behavior) to produce human-level performance in complex visuospatial tasks (playing Atari computer games; Mnih et al., 2015). The system only took raw pixel intensity values as input, after which it autonomously discovered complex series of strategies (goals and corresponding subgoals, e.g. taking elaborate detours through a maze) and action sequences (series of jumps and other complex movements) to maximize the outcome of the game (increasing the total score). Similar systems have since shocked the world by beating human experts in activities as diverse as media classification (Simonyan and Zisserman, 2014; Tran et al., 2015), medical diagnostics (Litjens et al., 2017) and the game of Go (Silver et al., 2017) and are quickly finding their way into robotics (Sünderhauf et al., 2018). In short, recent history shows that adding hierarchical structure to the various components of a control system has contributed much to their enormous success.

As illustrated above, the idea that living systems behave as hierarchical control systems is hardly new. Despite its firm rooting within the field of psychology and neuroscience, however, the concept of hierarchical control has been studied largely from the perspective of engineering and computer science, devoting little attention to the finer details of the architecture and function of living systems. Conversely, the idea that biological networks can be modeled as hierarchical control systems has escaped systematic attention in the biological sciences. In the past two decades, there has been a tremendous increase in our knowledge of the structure and function of living systems. This has shown that organisms follow generic rules of structure and function that apply universally to all living systems (see below). These insights have only partly been integrated with the field of control theory and machine learning. The purpose of the current paper is to bring these two influential fields of science further together. We will show that biological systems have a generic network structure that makes them ideally suited to function as hierarchical Bayesian control systems. Such systems can extract increasing amounts of contextual information from their inner and outer environments, construct increasingly articulated goal hierarchies and generate increasingly complex action sequences in order to reach (long-term) stability. We then identify a universal (stress)response of organisms to contextual cues that overtax their regulatory capacity and ability to remain stable. Such rules can be used to model organisms of any type, including humans.

## 3. Biological systems

Network science is booming, ever since the (re)discovery some 20 years ago of the *small world* network structure (Milgram, 1967; Watts and Strogatz, 1998) and the subsequent demonstration that universal laws of network theory govern network structures across a wide range of biological, psychological and social systems (Barabasi, 2013; Barabasi and Bonabeau, 2003; Barabasi and Oltvai, 2004; Barzel and Barabasi, 2013a,b; Newman et al., 2006; Oltvai and Barabasi, 2002). Because of its ability to connect different fields of science using a single methodology and corresponding terminology, network science holds considerable promise as a unifying discipline for many different fields, including biology, ethology, psychology and sociology. Below, we will first summarize some of the main findings from translational network science and identify generic rules of network architecture and function that apply to all living systems. We will then examine generic changes in biological systems when put under severe levels of stress.

### 3.1. On the structure of biological systems

At some level of abstraction, the whole of living nature can be considered to represent the interaction between building blocks that cluster together to form new building blocks, and so forth, until complex multicellular life develops (Oltvai and Barabasi, 2002). Collections of molecules form organelles, which in turn form cells, which in turn form tissues, which in turn form organs, organisms, organizations, biotopes, and so on. At each scale level of biological organization, the interaction between the building blocks that exist at this level (be they organelles, cells, organs, or organisms) can be visualized as a network structure in which building blocks are represented by nodes and their mutual connections by links. Almost without exception, biological networks show a topological structure called the *small world* structure, meaning that most nodes have few connections but some have many (the so-called *hubs*; Fig. 3). Hubs interconnect the various nodes of the network, allowing any two nodes in the network to be connected through a small number of intermediate steps, hence the term '*small world*' (e.g. all people in this world are an average of only 6 degrees of separation apart). Hubs contract large numbers of nodes into densely connected clusters (also called *communities* or *modules*; Girvan and Newman, 2002; Newman, 2006). The nodes that lie within such clusters share more connections amongst themselves than with other nodes within the network, forming subnetworks of their own. *Small world* structures are a general hallmark of biological systems and can be observed throughout living nature (Fig. 3).

*Small world* network structures turn out to be scalable, meaning that network clusters may themselves serve as nodes in a new network structure at a higher scale level of spatial organization, and so on. Thus, biological networks form hierarchies of part-whole relationships, in which higher levels of organization cannot exist without their constituent lower levels of organization (i.e. they form conditional dependencies in *space*) (Ravasz and Barabasi, 2003). Each new scale level again conforms to a *small world* network structure with multimodular features, which is why this architectural principle is called *scale-invariant*, or *scale-free* (Fig. 4; Barabasi, 2009). The scale-invariance of *small world* network structures has been compared to mathematical constructs called *fractals*: self-similar shapes that follow relatively simple algebraic rules across multiple scale levels of aggregation (Gallos et al., 2007;

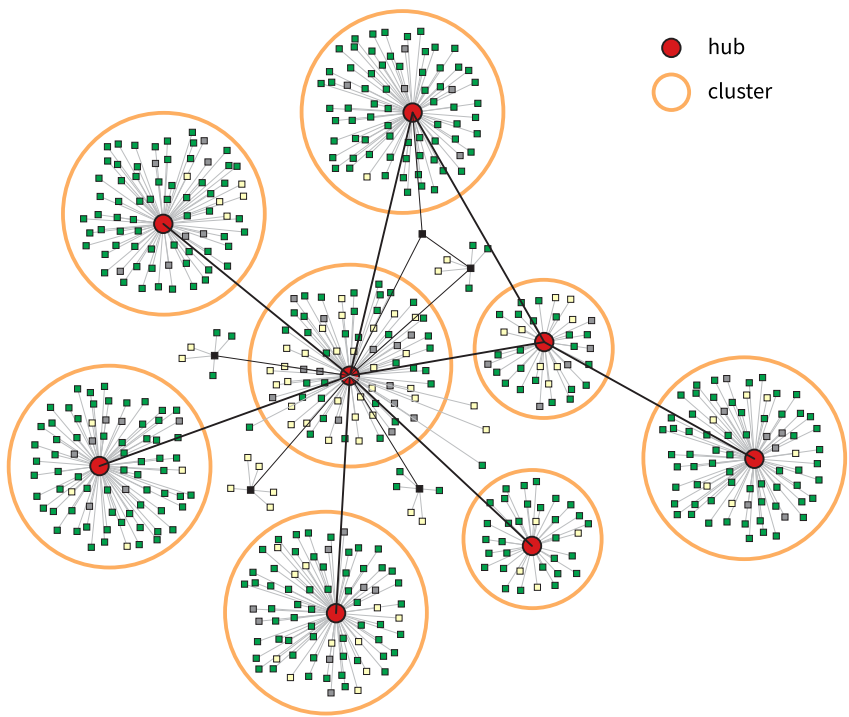

**Fig. 3.** Organisms as Small world Network Structures.
*Note:* Organisms can be conceived of as *small world* network structures. In such networks, hub nodes interconnect all other nodes in the network in such a way that the network as a whole has a small average pathlength (i.e. each node is only a small number of steps away from any other node in the network). In *small world networks,* hubs contract parts of the network into communities (modules), which are collections of nodes that share more connections amongst themselves than with other nodes. Because of such features, *small world* networks allow for highly efficient forms of information transfer at low wiring costs with a high tolerance for random damage. They are found in any 'connectome' studied thus far, including genomes, proteomes, metabolomes, microbiomes, neural connectomes, food webs and social networks.

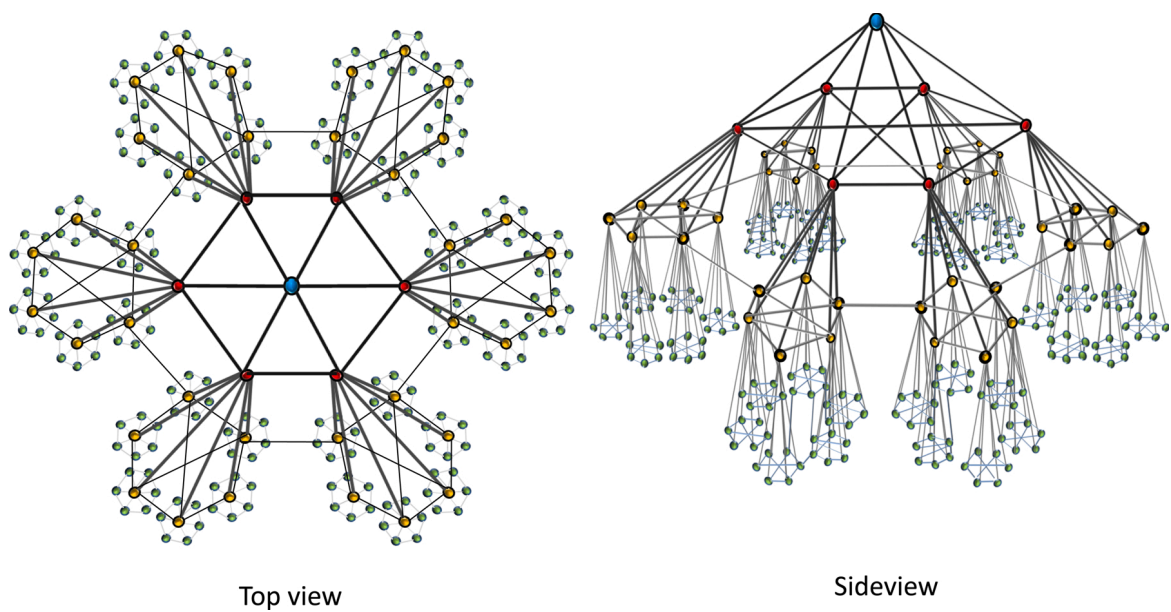

Top view　　　　　　　　　Sideview

**Fig. 4.** Scale Invariant Structure in Small world Networks.
*Note:* Schematic representation of scale invariant structure in *small world* networks. In such networks, hub nodes contract sets of other (hub) nodes into network clusters. Such clusters may themselves be considered nodes that cluster into superclusters and so on, producing a hierarchy of part-whole relationships. A *small world* network topology (see text) is found at each spatial scale level of biological organization, which is why this topological feature is called 'scale invariant' or 'scale free'. Blue node: central hub, which connects a set of 6 red nodes into a single network cluster. Red nodes are themselves hubs that each contract a set of 6 yellow nodes into another network cluster, etcetera. Note that this process of nested clustering can be repeated almost endlessly, illustrating the concept of scale invariance of *small world* network topology (i.e. any node in his figure may be a network cluster, supercluster, and so on; blue nodes may be drawn into clusters by high level hub nodes, or green nodes may become hubs by adding nodes). Right picture shows a sideview of the left image in which the vertical position of a node indicates its position within a nested hierarchy of hub nodes (a 'rich club'; Opsahl et al., 2008).

Song et al., 2005, 2006). Nested modular network structures such as these form spontaneously under the right conditions (i.e. a constant flux of energy into open dissipative systems), since such topologies allow network systems to get rid of their excess energy in the most efficient way, by minimizing resistance to energy flow (Jarman et al., 2017). A basic thermodynamic rule therefore suffices to produce network structures with short and efficient paths: a phenomenon called *self-organization* (Ashby, 1947; Kauffman, 1993). It has been hypothesized that life kick-started from *small world* networks of chemical reactions, which subsequently adapted to meet the more complex demands of life (Kauffman, 1996; Ramstead et al., 2018).

## 3.2. On the function of biological systems

Biological networks are not just static structures. Energy[1] flows through such structures in the form of information, e.g. chemical reactions at the level of receptors and genes, or electromagnetic changes at the level of neurons. In *small world* networks, some parts of the network receive energy (input) from the environment and change their states accordingly. These states are then altered as they flow on through the network in ways that depend on the wiring patterns of the nodes and modules in that part of the network (throughput). The processed states are then passed on to other nodes and modules (output) that lead to some action out into the environment. This succession of state changes is often referred to as network 'function'. Apart from universal rules of network structure, studies are now beginning to identify rules of network function that apply across different species and scale levels of biological organization (Barzel and Barabasi, 2013a,b; Friston, 2012; Gosak et al., 2018; Kitano, 2004). For instance, the input-throughput-output (I/T/O) organization of most biological networks turns out to resemble the shape of an hourglass, or 'bow-tie' (Fig. 5, left image; Csete and Doyle, 2004; Kitano, 2004). The input parts of these structures involve multiple input streams converging onto hub structures, which in turn converge onto higher level hub structures, etcetera, following a hierarchy of part–whole relationships in an upward manner. This goes on until a limited number of high-level hub structures is reached (i.e. the throughput parts). The output parts then involve multiple outputs diverging from these throughput hubs onto lower level hub structures and so on, down the nested hierarchy to the level of individual nodes (Fig. 5, right image). For example, a large number of sensory receptors and corresponding second messenger pathways fan in to a relatively small number of nuclear genes (the waist of the hour glass, or the knot of the bow-tie). Multiple outputs then fan out from these genes in the form of messenger RNAs that instruct a large number of ribosomes to produce all kinds of proteins that are cleaved into even more proteins (Barabasi and Oltvai, 2004; Watson et al., 2015; Zhao et al., 2006). A similar organization can be observed in the human brain (Markov et al., 2013). Here, a large number of neural columns within the visual cortex (coding for color, texture, speed, orientation, etcetera) converge onto a smaller number of brain areas involved in object representations, which in turn converge onto a few brain areas coding for global visuospatial scenes. This convergence goes on until anterior and frontal areas are reached that harbor some of the most global ('domain general') representations of the inner and outer environment (the waist of the hourglass). These global states then bias activity levels in several subordinate brain areas involved in the planning and execution of motor programs, which control a multitude of pyramidal cells and muscle fibers to produce motor action (Badcock et al., 2019; Bullmore and

Sporns, 2009; Freeman, 2005; Mesulam, 1998, 2008; Meunier et al., 2010). Bow-tie structures have been observed in the immune system, the internet and within other bow-ties (i.e. bow-ties nested within bow-ties), making this motif a scale invariant phenomenon (Box 1; Friedlander et al., 2015; Kitano, 2004; Zhao et al., 2006).

The ubiquity of the bow-tie motif has sparked questions regarding its functional significance. Bow-ties allow biological networks to convert a host of different inputs into a multitude of outputs using a minimal set of basic operations. Novel inputs and outputs can be easily plugged into a generic core of hub processes without affecting the system as a whole, making it a highly versatile structure. Thus, biological networks can combine robustness with adaptability in a chaotic world full of stimuli (Kitano, 2004). Simulation studies show that hierarchical networks spontaneously evolve bow-tie structure under some restrictions (Friedlander et al., 2015). Resources need to be scarce, and the evolutionary 'goal' that these networks aim to satisfy needs to be 'compressible', i.e. it should be possible to represent subordinate goal states by an increasingly small number of higher-level variables without losing too much information. This continues until the top of the hierarchy is reached (the knot of the bow-tie, or the waist of the hourglass). The minimal width of the bow-tie structure therefore represents the maximum level of compression of an evolutionary goal, with subordinate structures representing lesser compressed versions of the goal state (Friedlander et al., 2015). As we shall see below, this aspect of bow-ties structures turns out to be rather fundamental: a high-dimensional input is forced through a bottleneck, or low-dimensional manifold. This relates to the concept of dimensionality reduction which can be found throughout statistics and machine learning (e.g. principal component analysis and other clustering methods (Sorzano et al., 2014)). Studies have shown that imposing an 'information bottleneck' structure onto hierarchical (deep) networks significantly increases their performance by allowing for some form of compression and generalization of events that take place at lower levels (Hafez-Kolahi and Kasaei, 2019; Shwartz-Ziv and Tishby, 2017). Apparently, living systems minimize complexity cost and use the fewest degrees of freedom to model their environments, i.e. Occam's principle (Maisto et al., 2015). Organisms can therefore be conceived of as dimension reduction machines that perform a hierarchical clustering on input in an attempt to find the most parsimonious (global) representation without losing too much information. Such high-level compressed representations then fan out to the lower parts of the output hierarchy to produce coordinated action sequences. In short, the bow tie motif provides organisms with an optimal infrastructure to function as hierarchically organized (and model based) control systems.

The flow of information across bow-tie network structures is not a simple process with energy flowing directly from input (via throughput) to output areas in a linear fashion (Kitano, 2004). Bow-tie structures may show cross-connections (shortcuts) between their input and output parts at different levels of the hierarchy, causing the structure to fold back onto itself (Fig. 5, right image). This produces short input–throughput-output loops near the bottom of the hierarchy as well as longer loops that run from input to output along progressively longer throughput loops, reflecting different degrees of processing (Fig. 6). Additionally, feedforward and feedback loops run down and up the hierarchy respectively, reflecting predictive coding as well as corrections of such predictions by means of novel input (Box 1, Fig. 6). Such structures differ from hierarchical control systems that are traditionally used in engineering and machine learning and come with specific functionality. In recent years, insights have grown that organisms are not merely reactive agents that respond passively to external stimuli. Rather, they seem to actively model the causal structure of their inner and outer worlds and use memories to predict future events in a biological equivalent of Bayesian inference (Box 1). The idea that biological organisms engage actively in some form of hierarchical Bayesian inference has produced an explosion of literature in the past decade (Box 2). In this view, each level within a hierarchy generates a predictive model of the hidden causes of the effects (events, activity) observed at a lower

---

[1] We will use the notion of energy to stand in for the dynamics that couple different nodes or clusters in network graphs. Technically, the energy can be thought of as a log probability of a given state of a node or cluster (i.e. the rarity of a given state, which serves as measure of information content) and the dynamics of network systems can usually be framed in terms of gradient flows on this log probability (i.e. gradient flows on rarity, or information content). A nice example is given by the free energy principle (see below), which defines prediction error as the gradient of variational free energy. In other words, when we talk about energy flows (and network function) we are actually talking about gradient flows on free energy that usually have an interpretation in terms of information flows. As we will see below, this corresponds to predictive coding and Bayesian belief updating in a variational setting (e.g. in artificial or biological systems).

[2] In this paper we refer to (anatomically) backward or descending connections as (control theoretic) feedforward connections, which convey predictions ('Bayesian beliefs'). Conversely, we refer to (anatomically) forward or ascending connections as (control theoretic) feedback connections, which perform an update on predictions after measurement (this is called 'Bayesian belief updating', see below).

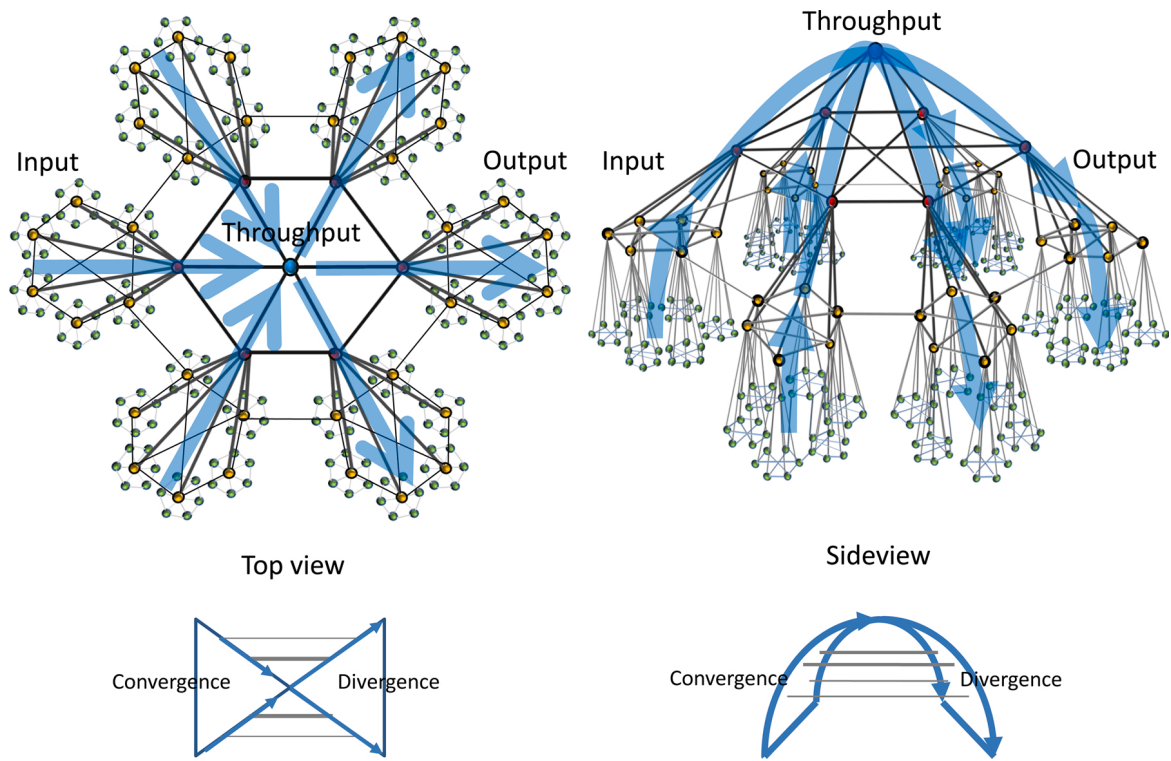

**Fig. 5.** Organisms as Nested Modular Small World Networks: The Bow-tie (Hourglass) Motif.

*Note:* **Left image:** organisms can be conceived of as nested modular *small world* network structures with a distinct input-throughput-output organization: a bow-tie (2D) or 'hourglass' (3D) structure. The input parts of such networks involve multiple energy streams converging onto each other while ascending in a hierarchy of part-whole relationships (left part of bow-tie). Conversely, the output parts involve multiple energy streams diverging while descending in the hierarchy (right part of bow-tie). The 'knot' of the bow-tie (or the waist of the hourglass) lies in between its input and output parts (i.e. the throughput part). This motif can be observed across multiple scale levels of organization, making it a scale invariant feature. **Right image:** bow-tie motifs may show cross-connections (shortcuts) between their input and output parts at different levels within the hierarchy, causing the structure to fold back onto itself (right figure). This allows energy to travel from input and throughput to output structures across loops of various pathlengths, corresponding to different degrees of information processing (see Fig. 6). Please note that arrows in this figure only show the global direction of energy flow. Feedback and feedforward connections[2] run up and down the various levels of the hierarchy, which are thought to represent prediction errors and predictions relative to lower-level input (Box 1).

hierarchical level of organization. The error between the model and the organism's sensory states served both to update the model and to inspire action, which changes the environment to alter perception, after which the process reiterates. Such generative models perform well across a limited number of observations and trials and their predictions generalize well beyond the subset of training data, suggesting that a certain amount of 'creativity' is involved in hierarchical Bayesian modeling (Blei et al., 2017; Tenenbaum et al., 2011). This creative property of higher level models has been linked to the concept of emergence in complex systems (Griffiths et al., 2010; McClelland et al., 2010). So far, however, it has remained largely unclear how this type of processing is implemented in biological systems. This will be discussed in the next section.

### 3.3. Organisms: nested modular network structures that function as hierarchical Bayesian control systems

In the previous section, we saw that organisms are optimally wired to function as hierarchical (nested modular) control systems that combine hierarchical perception (input) with hierarchical goal setting (throughput) and hierarchical action control (output), to iteratively respond to their environments. Below, we will discuss how energy flows travel through such systems (hierarchical message passing) to support Bayesian inference, turning organisms into hierarchical Bayesian control systems.

Organisms do not simply respond randomly to environmental stimuli. Rather, they must connect input patterns to output patterns in ways that are compatible with life, e.g. when the input is 'food' (glucose), a suitable output would be 'approach'. When the input is 'predator' (smell), a suitable response would be 'avoid'. Such non-random responses are called 'adaptive', since they allow organisms to adapt to changing environmental conditions and survive (Gross and Blasius, 2007). Connecting input patterns to adaptive output patterns ('policy selection') can be a daunting task for any organism, however. Most organisms live in a rich context of environmental circumstances, which contains multiple cues that may elicit conflicting responses (e.g. approaching food, but avoiding a predator). Such conflicts must be resolved in order to survive (i.e. responses must be prioritized and put in sequence). This requires organisms to build an integrated rather than segregated representation of their environments (e.g. input (food| predator), instead of input (food), input (predator)). Because of their peculiar structure, nested modular (hierarchical) network structures are optimally suited to produce such integrated representations (van den Heuvel et al., 2012v). The input parts of such structures involve multiple inputs that converge onto fewer hub structures (Fig. 5). Like spiders in a web, such hubs keep in touch with the states of large numbers of functionally segregated nodes and clusters in the network, each of which confers part of the relevant information concerning the inner and outer environment. The state of such hubs thus provides a summary representation of the states of all nodes that connect to it (i.e. a state with a higher level of parsimony and abstraction than its subordinate substates, e.g. input (food|predator)). Such functional integration goes on until the top of the nested hierarchy of network clusters has been reached. At each level within the input hierarchy, integrated input states are compared to integrated reference states at a similar hierarchical level (e.g. throughput (food|predator), after which the ensuing errors are

---

**Box 1**

**On the Structure of Organisms: Network Motifs and Predictive Modeling**

Biological (*small world*) networks are made up of smaller building blocks ('subgraphs') with a relatively large scale called 'network motifs'. These are highly generic pieces of network structure that are observed across different spatial scale levels of biological organization, where they support similar functions (e.g. speeding up or slowing down responses, prolonging responses, integrating or coordinating states, etcetera). The bow-tie structure is just one of these building blocks, with a relatively large size. When examining their finer substructure, bow-ties consist of a family of smaller motifs (Alon, 2007; Araujo & Liotta, 2018; Li et al., 2012). Studies have found a particular abundance of the so called 'feedforward loop' (FFL) in living systems (Alon, 2007). This is a motif that consists of only three nodes (A, B, C) with directed connections between them (i.e. A=>B, B=>C and A=>C). Typically, FFLs lack a connection that runs from the output of the motif back to its input (i.e. C=>A), i.e. they are open loop control systems. When confronted with a stimulus, such motifs push forward a 'best guess' response regardless of its outcome, hence the term 'feedforward'. Because of their ability to forward best guess responses, feedforward motifs have been linked to predictive processing (Del Giudice et al., 2018). For instance, the act of eating already increases insulin secretion regardless of actual increase in blood glucose concentration, which involves a predictive feedforward system (Marchetti et al., 2008; Pezzulo et al., 2015; Pezzulo et al., 2018). In contrast, feedback motifs contain links that run from the output nodes back to the input nodes, i.e. they are closed-loop control systems. Such systems represent events that are the consequences of certain actions. For example, feedback systems are involved in measuring actual blood glucose concentrations after eating, to provide an update on the predictions made by feedforward motifs (Marchetti et al., 2008). The activity of feedforward and feedback systems needs to be balanced in order to have the best of both worlds. In biological systems, FFL motifs represent the feedforward (descending) propagation of predictions from higher levels to lower levels within the nested hierarchy, whereas feedback motifs represent the prediction error that runs back in the opposite (ascending) direction[2]. Thus, feedback and feedforward loops run up and down the bow-tie hierarchy, respectively, to balance prediction errors with predictions. This balance is what underlies 'hierarchical predictive coding' in living systems (see next section, Fig. 6). The ubiquity of FFLs in living systems suggests that predictive activity makes up a substantial part of these projections. This corresponds to cumulative findings that organisms are not merely reactive agents but rather proactive and 'predictive' agents that use memories to predict future events. One of the best known examples is anticipatory salivation in classically conditioned dogs, but Pavlovian learning and anticipatory responses such as these have been demonstrated in organisms as simple as bacteria (Brembs, 2003; Calvo & Friston, 2017; Friston, 2012; Hesp et al., 2019; Mitchell et al., 2009; Tagkopoulos et al., 2008). In short, the nested hierarchical bottleneck structure of bow-tie motifs and their constituent submotifs are a scale free feature, which provides living systems with an optimal infrastructure to function as hierarchical control systems at every scale level of their anatomy (Friston, 2012; Hesp et al., 2019; Ramstead et al., 2018).

---

conferred to output hubs at a similar level, which then disseminate across lower levels of output hubs, etcetera, to eventually generate complex output sequences (see Section 2).

While encoding their environments, organisms have to solve the 'binding problem' i.e. they need to decide whether signals come from a common cause and should be integrated (i.e. represented by a single node or cluster), or whether they come from independent sources and should be segregated (i.e. represented by separate clusters) (Rohe et al., 2019). The number of independent clusters in a nested modular hierarchy is therefore a function of the number of independent contextual cues that need to be controlled by the organism (Ashby, 1956, Conant and Asby, 1970; Edlund et al., 2011). When environments grow more complex, organisms need to incorporate more clusters in order to produce increasing articulated (contextualized) models. Functional integration across these clusters then increases the hierarchical depth of such systems, allowing for increasingly abstract representations. For instance, some organism can be confronted with food (input A), a mate (input B), a rival (input C) and a predator (input D), all at the same time. It then has to choose whether to eat (output 1), mate (output 2), fight (output 3) or hide (output 4), given its own internal state hungry (input a), alone (input b), wounded (input c), or weak (input d). Each of these factors needs to be encoded into a separate node or cluster ('functional segregation', 'specialization' or 'factorization', e.g. input (A, B, C, D, a, b, c, d)). The functional integration across such perception primitives then produces a hierarchy of part-whole relationships that allows for increasingly abstract (contextualized) percepts when moving up the hierarchy (e.g. input(A|B|C|D|a|b|c|d = input(X)). Similarly, adding primitives to the output parts of a bow tie allows for a richer repertoire of actions (e.g. output(1, 2, 3, 4)). A deeper integration across such clusters produces more elaborate forms of action control and more complex forms of behavior (e.g. 'courtship', which may involve complex action sequences (e.g. output (3|1|4|2) = output (X)). However, extending the repertoire of input-output strategies raises chances that such policies will conflict with one another. In living systems, these conflicts are resolved in a hierarchical fashion (e.g. throughput (A|a) → output(1), throughput (A|B|b) → output(2), throughput (A|B|C|a|b) →

output(3), throughput (A|B|C|D|a|b|c|d) → output(4)). More complex environments therefore require organisms to not only expand their input and output hierarchies, but also their throughput hierarchies, in order to connect input to output strategies in a non-random (adaptive) manner for different combinations of events (i.e. policy selection). In other words, organisms develop hierarchies of reference states and corresponding substates, which are called 'goal hierarchies' (Pezzulo et al., 2015, 2018).

Apart from resolving conflicts between opposing policies in (current) space, goal hierarchies are used to solve potential conflicts in time. For instance, my *current* input state input$_t$(A|B|C) (= being warm, well fed, no predators) seems to match my current goal state throughput$_t$(A|B|C) and output pattern output$_t$(lying down), but this policy may well conflict with my *anticipated* input state throughput$_{t+1}$(D|E|F) (e.g. being cold, hungry, lurking predators) and corresponding output output$_{t+1}$(heating, eating, locomotion) (De Kleijn et al., 2014). To resolve 'temporal' conflicts, the same principle of hierarchical control that allows organisms to integrate increasing numbers of contextual cues in space can be used to integrate contextual cues in time: temporally more distant goal states are encoded by control systems that are superposed onto those that predict temporally more proximal ones in a hierarchy of part-whole relationships (Pezzulo et al., 2018). Errors that are produced relative to such predictive goal states ('prediction errors') may result in actions at a time when such events have not yet taken place (e.g. foraging, stacking fat, storing food, finding shelter, building nests, feeding offspring, preparing to attack). This involves a time and energy investment that is not immediately contingent to the current situation, but serves to keep the system stable through change (i.e. 'allostasis', see introduction). Thus, the ability of organisms to predict events at least some time ahead allows them to engage in 'pre-emptive' actions that significantly raise their chances of survival. The act of anticipating increasingly complex events ever more distantly into the future requires ever deeper hierarchies of goal states, which integrate across multiple levels of subgoals and corresponding timeframes to infer ever more global goal states (Pezzulo et al., 2015). Such highly integrated and predictive goal states are often referred to as 'world models', since they may involve quite complex anticipatory designs ('simulations') of the inner and outer environment of the organism, which inspire complex forms of behavior of an increasingly anticipatory nature (see Section 2 and below). Of course, not all organisms equally express goal hierarchies or world models. The height of such hierarchies varies from 'lower' to 'higher' organisms and between organisms of the same species, causing their behavior to vary along with it.

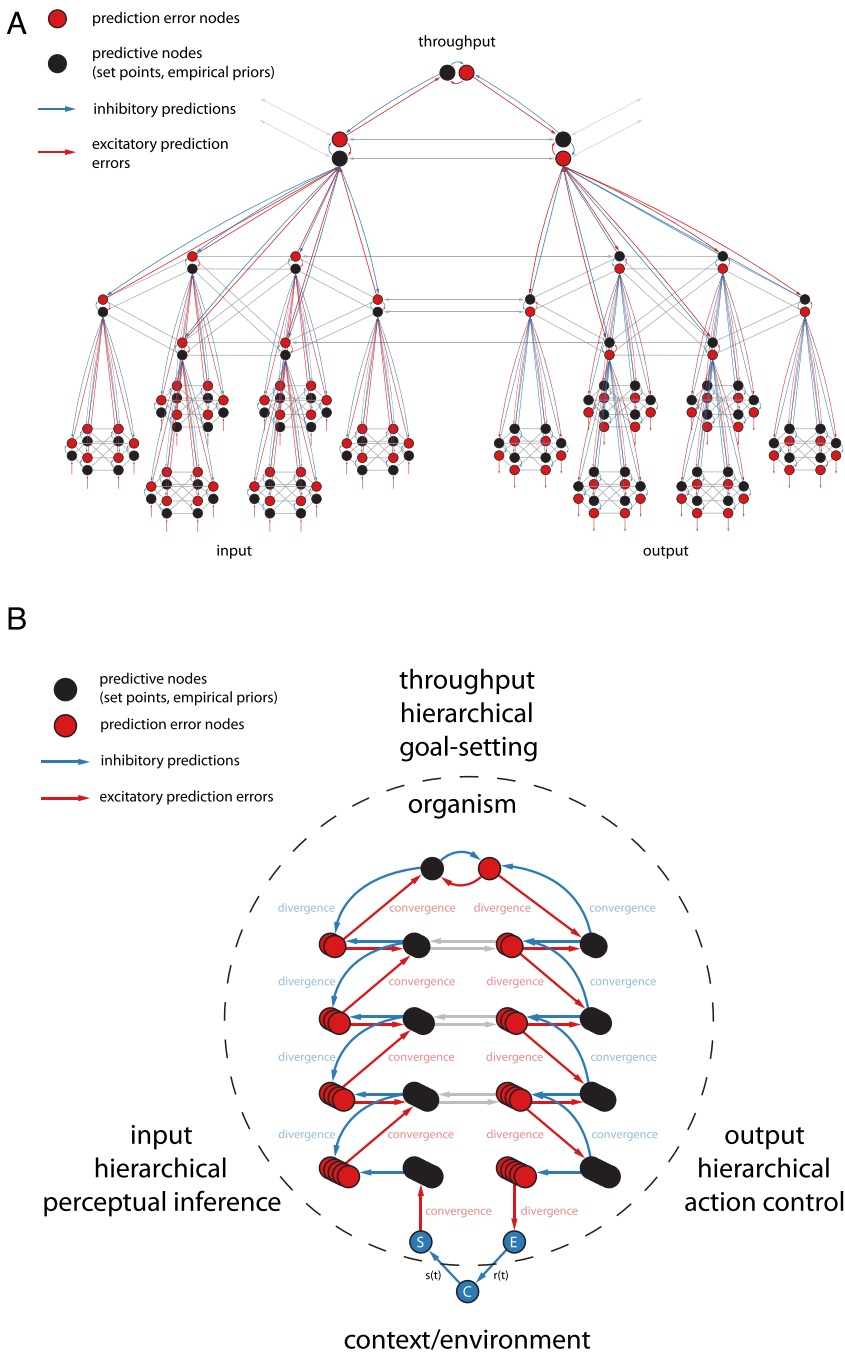

Rather than generating a predictive model for every possible contingency, organisms use memories to predict future events (e.g. the ringing of a bell causes anticipatory salivation in classically conditioned

**Box 2**

On the Function of Organisms: Active Inference and the Free Energy Principle

According to the free energy principle, the dynamics of biological systems follows from the basic laws of thermodynamics, i.e. organisms must find their lowest possible energy state despite a continuous influx of energy. In this view, living systems are statistical engines that encode models of the world simply by responding to their input (Friston et al., 2013). The difference between the actual input to the system and some predictive model of the world corresponds to the prediction error of the system, which under some restrictions corresponds to an information theoretic quantity called 'variational free energy'. Low prediction error corresponds to a low number of alternative states that an organism occupies on average and, therefore, a more stable, low-energy state that has been equated to 'homeostasis' (Friston, 2012). Suppressing prediction error is therefore an imperative for all living systems, since it amounts to finding a stable low-energy state. Organisms generally strive towards this overarching goal by generating world models with multiple levels of model complexity and by testing these models against incoming input by performing actions ('active inference'). Such actions change the environment of the organisms, which produces a novel input that is used as a test on model evidence. In other words, organisms act to maximize sensory evidence for their own predictions: they are 'self-fulfilling prophecies'. Organisms cannot only reduce prediction errors by changing the environment through action in order to alter their percepts ('changing your actions', as in perceptual control), but also by updating their world models to produce a better fit with their input states ('changing your mind'): a process called 'Bayesian belief updating' See text for further details. Although originally formulated within the context of human brain function, the active inference principle has been generalized to involve living systems across multiple spatiotemporal scale levels of organization, varying from microbes and brains to social systems (Ramstead et al., 2018). According to active inference theory, organisms 'are' embodied and situationally embedded (Bayesian) models of the world and natural selection is nature's way of performing Bayesian model selection (Hesp et al., 2019). For equations describing the free energy principle and the process of active inference, see (Friston, 2010, 2012) dogs). This allows them to restrict predictive modeling to (combinations of) events that have some probability of actually occurring (since they occurred in the past). The act of prediction is therefore intimately tied to the process of learning. Goal hierarchies connect input patterns to output patterns by means of non-random (adaptive) connections. The act of making non-random connections between the input and output of a controls system is called 'associative learning'. This involves the selective strengthening and weakening of connections within throughput areas, which may e.g. involve molecular bonds in signaling networks or synaptic connections in neural networks. Goal hierarchies develop in the course of an individual's life, as well as in the course of evolution: any failure to connect stimuli with adaptive responses during the course of their lives (ontogenetic learning) will cause organisms to be eliminated through natural selection (phylogenetic learning). The rewiring of different parts of such hierarchies has been linked to different types of associative learning (Pezzulo et al., 2015, 2018). Short stimulus-response loops represent simple autonomous and/or motor arc reflexes that allow for basic Pavlovian (stimulus–stimulus) learning and instinctive behavior. Longer loops allow for more complex forms of learning such as habit learning and corresponding behavior, whereas the longest loops involve true goal-directed learning and the formation of explicit world models that inspire goal-directed behavior. Goal hierarchies thus consist of progressively longer loops that run from input to output via different levels of integration within the throughput hierarchy. The various forms of associative learning that take place within goal hierarchies are thought to be universal to organisms at any scale level, with more (spatially and temporally) integrated forms of learning occurring within increasingly 'higher' organisms. Pavlovian (predictive) learning has been observed to occur within organisms as primitive as bacteria (Calvo and Baluška, 2015), whereas goal-directed learning is observed within higher vertebrates and some invertebrates (Pezzulo, 2012).

Although it is now increasingly recognized that organisms are predictive agents (see Box 1), it remains unclear how exactly predictive modeling is implemented in living systems. Having a nested modular network structure with information bottlenecks motifs appears to be a necessary precondition, but it is not a sufficient one. The graphical model of Fig. 4 therefore requires modification to allow for predictive coding. To this end, tentative hypotheses have been put forward that are based on hierarchical message passing in the human brain (Adams et al., 2013; Friston, 2018, 2019b; Friston et al., 2017; Kanai et al., 2015; Kiebel and Friston, 2009). In Fig. 6, we show the putative wiring scheme for hierarchical predictive coding in biological systems (Adams et al.,

2013; Friston, 2018; Kanai et al., 2015), which we adapted to accommodate a folded information bottleneck structure (a 'bow-tie' motif). Here, predictive states are encoded by nodes at a higher level of integration, which suppress prediction errors at lower levels of integration by means of divergent (disynaptic) inhibitory connections (Fig. 6). The difference (prediction error) is conveyed horizontally to the output hierarchy as well as projected back upward by convergent excitatory connections to correct these higher-level predictions (update the models), turning them into posterior expectations ('empirical priors'). This process is called 'Bayesian belief updating' and involves the actual learning process (i.e. a change in connective efficacy). Thus, higher level models attempt to suppress ('explain away') prediction errors produced by lower-level systems, whereas lower-level systems in turn correct higher-level predictions. Such circularly causal relationships produce oscillations that are typically observed in neural dynamics. For an overview of the mathematics describing the process of hierarchical message passing in the context of Bayesian inference, see (Kiebel and Friston, 2009).

The output hierarchy shows a similar but inverted makeup in exactly the same way described under the equilibrium setpoint hypothesis, or indeed perceptual control theory (Adams et al., 2013; Friston, 2019b). Here, prediction errors descend down the hierarchy while diverging onto lower-level hub nodes to correct low-level predictive models, whereas predictions ascend up the hierarchy while converging onto higher-level prediction error units. Thus, prediction errors globally ascend and converge within the input hierarchy and descend and diverge within the output hierarchy, to eventually supply the setpoints of lower-level output primitives (e.g. motor or autonomous reflex arcs). Each level within the input hierarchy tries to explain away prediction errors produced at lower levels within the hierarchy by means of inhibitory (predictive) connections (Fig. 6). If prediction errors cannot be suppressed by a simple (less integrated) world model and corresponding output produced at the bottom of the hierarchy, they are carried up to the next level in an attempt to suppress the errors using a more elaborate (contextually more integrated) model (see section 3.5). In action control, this process of hierarchical message passing takes place in inverted order. Here, prediction errors that have not been successfully explained away run down the hierarchy to inspire action. Such output may still reduce prediction errors within the input hierarchy by changing the environment and, hence, the input to the system ('active inference').

The process of predictive coding and belief updating as described above is thought to reflect hierarchical Bayesian inference in biological systems, and can be seen as a general model for information processing. It is thought that similar principles apply in any organism, from microbe to man (Friston, 2012, 2018; Hesp et al., 2019; Ramstead et al., 2018). For instance, membrane receptors and second messenger pathways may represent posterior expectations that are informed by genetic or biochemical priors (setpoints) at different levels to produce output. Such systems may produce oscillatory dynamics similar to those observed in neural dynamics (Friston, 2012). As can be seen in Fig. 6, information bottleneck motifs can be observed at the level of individual nodes, clusters, superclusters and within the network structure at large, i.e. it is a scale invariant feature. As a consequence, the organism itself can be modeled as one giant feedforward loop motif (Box 1), which produces predictive output that feeds back into the organism through the environment, providing an update on predictions ('active inference', Box 2). This means that Fig. 1 should be adapted to contain an arrow running from node A directly to node C.

At this point, it is important to emphasize the difference between traditional notions of hierarchical Bayesian inference in statistics and hierarchical inference as it takes place in living systems. First, statistical models usually involve a single hierarchical generative model. In living systems, the architecture of generative models acquires two streams: a sensory or input stream that controls input while is primarily concerned with inferring "what the world is doing" and an executive or output stream that tries to infer "what the organism is doing" (either in terms of motor behavior or autonomic function): the dual hierarchy in Fig. 6. Input hierarchies are involved in hierarchical perceptual inference, i.e. producing increasingly comprehensive perceptual models that try to explain lower-level sensory events (Friston et al., 2006). Output hierarchies on the other hand are involved in hierarchical action control, i.e. decoding high-level abstract models within the information bottlenecks of organisms into detailed action sequences produced by action primitives located at the base of the output hierarchy. Unexplained (residual) prediction error thus moves down the hierarchy to eventually supply the setpoints of low-level action primitives and pose as complex 'output commands'. Meanwhile, predictions with respect to the hidden causes of sensory events that take place in motor (e.g. proprioceptive) or endocrine (e.g. interoceptive) structures run upward in this hierarchy, in an attempt to suppress prediction errors. This counterstream represents feedback on the correct execution of motor or endocrine actions, based on the organism's models of what it is doing (i.e. based on the inferred sensory states of its output organs). When predictions with respect to the actual state of output organs (represented by bottom-up predictions) matches the output command (by top-down prediction error), prediction errors are fully suppressed and the execution of the output pattern comes to a halt. (Friston, 2019b). This dual aspect of hierarchical inference is emphasized by referring to nested hierarchical bow-tie network architectures (with small-world characteristics). This means that "bow-tie" should be read as a dual-aspect spatial hierarchy responsible for making inferences both about hidden states of the world and actions upon those states, respectively.

Second, models of hierarchical (Bayesian) inference in statistics are unfamiliar with the concept of goal-directedness (agency). This concept is still a topic of debate (Walsh, 2015), yet seems to be clearly definable from the perspective of organisms as hierarchical control systems. As observed in Section 2, perceptual control theory already equated the reference signal (setpoint) of control systems with goal-directedness and the hierarchical organization of reference signals with the formation of more complex goal states (Powers, 1973b). Similarly, model-based control theory involves organisms constructing elaborate hierarchical models of the world that serve as predictive goal-states that are encoded by intermediate throughput areas (Solway and Botvinick, 2012). In active inference theory, goal states align with so called empirical priors. These are nodes or clusters that encode prior beliefs that have been updated by sensory input, i.e. priors at intermediate levels within a hierarchical model (the black nodes and clusters in Fig. 6). Such nodes or clusters encode the states, or sensory information sampled, that the organism *a priori* prefers to occupy or sample, after having been updated by a certain input (red nodes in Fig. 6). Goal states can therefore be construed as 'posterior expectations and beliefs about controllable but hidden states of the world'. The scale free nature of living systems makes sure that empirical priors form nested hierarchies, with higher-level clusters of priors reflecting increasing amounts of contextual integration of preferential or predictive states (i.e. from individual setpoints to complex world models). In other words, the nested modular hierarchy of black nodes and clusters in Fig. 6 (empirical priors) reflects a hierarchy of goals and corresponding subgoals, down to the level of individual setpoints. Similarly, the nested modular hierarchy of red nodes and clusters in Fig. 6 (prediction error units) represents a hierarchy of empirical evidence at different levels of contextual integration, which is aligned along the various levels of the goal hierarchy to provide an update on these models. Thus, instead of being fixed and given, goal states are progressively inferred within the narrowing bottlenecks of bow-tie structures that form a smooth continuum between input- and output hierarchies (Fig. 6A). These structures are involved in inferring "what the organism should be doing", i.e. hierarchical goal-setting. As a result, we necessarily introduce the notion of 'hierarchical Bayesian control systems'. Such systems combine hierarchical perceptual inference (input) with hierarchical goal inference (throughput) and hierarchical action control (output), to eventually reduce overall levels prediction error through active niche exploration ('active inference').

This concludes our description of how goal hierarchies are constructed in living control systems. Below, we will examine which global types of goal states are produced within deep goal hierarchies and discuss their putative positions within a nested hierarchy of network clusters. We will then show how such hierarchies collapse in a top-down manner under rising levels of stress, leading to corresponding changes in behavior.

### 3.4. A taxonomy of goal states

The central tenet of hierarchical Bayesian inference in biological networks is that organisms try to infer the hidden causes of their sensory input (effects) and construct predictive models to do so. The difference (error) between these predictions and the perceived events is used simultaneously to inform behavior (output) and to adjust the model (Friston, 2012). From both observational data and theoretical considerations, organisms are thought to construct at least two global types of predictive models (goal states) at the top of their goal hierarchies. One of these involves a model of the organism itself (Limanowski and Blankenburg, 2013; Moutoussis et al., 2014). Since any organism has a body, it will consistently receive input that can be explained as produced by or originating from within that body. Such signals may involve both changes in the internal state of the organism (e.g. changes in internal glucose or acidity levels) as well as changes in its external environment as a result of actions produced by the organism itself (e.g. chemicals secreted or vibrations produced by the organism itself). Through hierarchical Bayesian modeling, organisms will eventually infer the hidden common causes behind these various types of signals (effects) and, eventually, the 'self' as a single common cause. Prediction errors relative to such 'self-models' produce behavior that shows hints of a sense of agency (e.g. a differential response to signals produced by the organism itself rather than its environment). The principle of hierarchical Bayesian inference therefore predicts that self-models are produced to varying degrees in any organism, from microbe to man. Most organisms have different sensory systems for monitoring their inner and outer worlds. 'Exterocepsis' is used for sensing external events and usually includes 'the 5 senses', i.e. vision, smell, hearing, touch and taste. Interocepsis is used to monitor internal events and involves sensory streams from smooth muscles, endocrine glands and other organs. Proprioceptis is used to relay the state of the world in between the interior and the exterior of the organism and involves data streams from striated muscles (e.g. muscle spindles). Internal/self-models rely disproportionally on sensory information derived from internal systems (e.g. interocepsis and proprioception) (Limanowski and Blankenburg, 2013; Moutoussis et al., 2014). Since output hierarchies are primarily involved in inferring 'what the organism is doing' and rely disproportionally on internal sensory streams to do so, internal/self models likely form a continuum with the top of the output hierarchies of bow-tie structures (Figs. 6 and 7). This at least seems to be the case in the human brain (Seth and Friston, 2016; Smith et al., 2019a; Thornton et al., 2019).

Similarly, organisms can infer the (hidden) common causes of effects (input) produced by factors outside of their own body (Baker et al., 2017; Limanowski and Blankenburg, 2013; Ondobaka et al., 2017). Apart from abiotic factors such as rain or snow, such external factors may involve models of other organisms and their intentions (e.g. predator, prey, friend or rival). Such social models are produced to varying degrees in any organism. Prediction errors relative to these models inspire social behavior, which reflects some sense or knowledge of the agency of other organisms, i.e. their existence, social roles, needs and intentions. Such behavior can be found already at the level of bacteria (e.g. quorum sensing in biofilms; Lyon, 2015). External/social models predominantly rely on sensory information derived from input organs (e.g. exterocepsis) to derive the state of the external world (Smith et al., 2019b; Moutoussis et al., 2014). Such information is then passed on to internal systems to formulate an adaptive response and monitor its execution. Meanwhile, external/social models control the output of the same external systems in a hierarchical manner (i.e. attentional biasing of exteroceptive organs). Since input hierarchies are concerned with inferring what the external world is doing, it is therefore likely that external/social systems form a continuum with the top of the input hierarchies of bow-tie structures whilst being strongly connected to internal/self systems (Fig. 7). Again, this at least seems to be the case in the human brain (Seth and Friston, 2016; Smith et al., 2019a; Thornton et al., 2019).

As observed in section 3.3, the complexity of a goal hierarchy may vary across individuals and species depending on environmental complexity, and the behavior of their owners varies along with it. We therefore predict that organisms that display a greater degree of agency should show a local extension of their nested hierarchical trees to encode more explicit self-models, i.e. involve the integration across a larger number of network communities. This hypothesis can be tested e.g. by examining organisms that differ in the degree to which they respond differentially to (chemical or physical) signals produced by themselves rather than their environment, or the degree to which they show signs of (self-referential and goal-directed) behavior (agency). Such organisms should have larger scores on measures of hierarchical depth within specific parts of their networks (see Discussion). Similarly, we propose that social behavior, when compared to solitary behavior, should involve some local extension of their hierarchical trees to encode more explicit social models. Such models may become especially intricate in highly sociable species that spend a lot of time gauging the social roles and intentions of their community members (e.g. some birds, mammals and primates). Such organisms are constructing world models of the world models of other organisms (i.e. recursion and reciprocity; Friston and Frith, 2015). These hypotheses can be tested by comparing the hierarchical network structure of solitary and social species, or social species that differ in their level of sociability (see Discussion). A similar argument can be made for the ability to predict events ever more distantly into the future. We predict that temporally more distant goal states require deeper hierarchies of control i.e. the integration across a larger number of network communities. This can be seen as a hierarchical extension of interior (self-referential) and/or exterior (social) models to accommodate long-term predictions with respect to self and/or others. Such anticipatory actions may be aimed at a future version of the individual itself or some external agent, rather than the current self or the current other. This hypothesis can be tested by comparing the hierarchical depth of network (bow-tie) structure between individuals or species that differ in their ability to anticipate (self-referential and social) events (see Discussion).

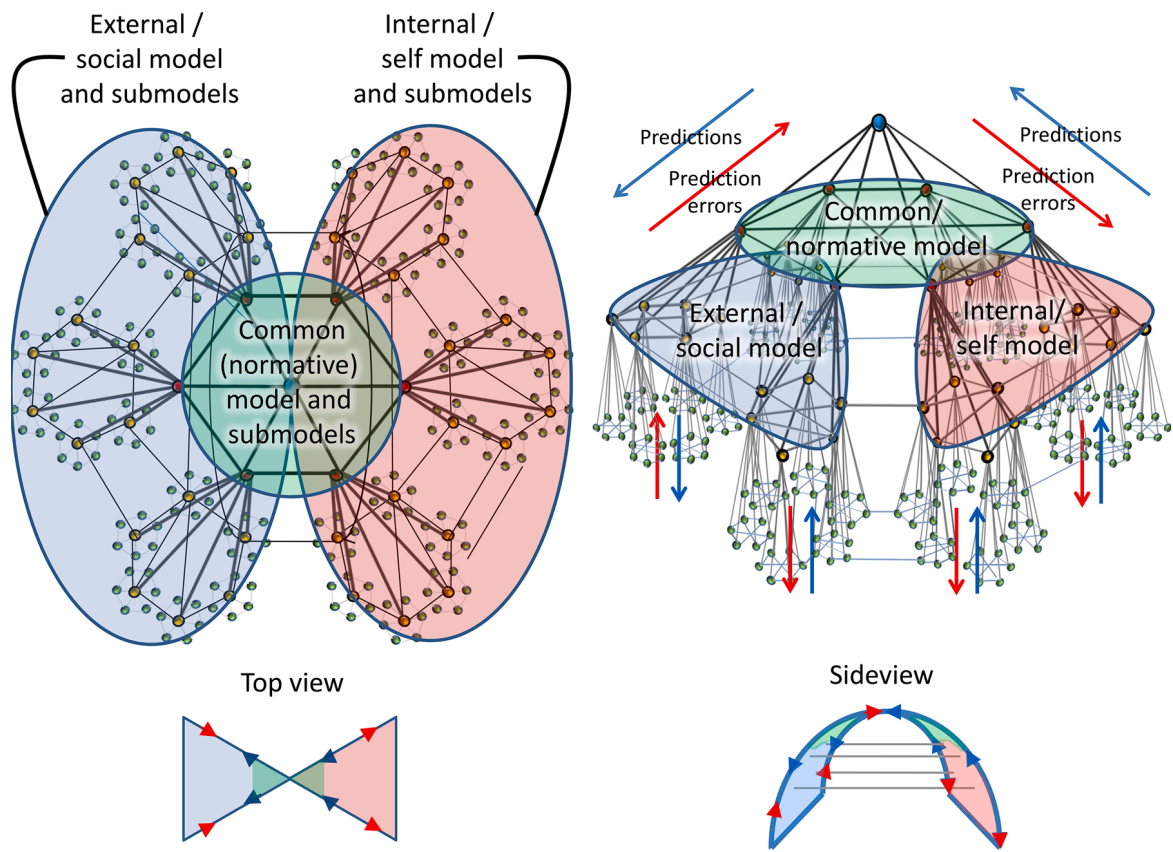

**Fig. 7.** Putative Relative Positions of High-level Goal States Within Living Network Systems.
*Note:* Schematic view of the way in which higher level ('normative') world models may develop within a hierarchy of goal states through the functional integration across self and social models. External (social) models predominantly involve inference on exterocepsis and may hence form a continuum with the input-part of a goal hierarchy. Internal (self) models predominantly involve inference of interocepsis and proprioception and may hence form a continuum with the output part of a goal hierarchy. Logically, cross-cutting (normative) models that integrate across internal and external world models and corresponding time domains form the top of the goal-hierarchy (i.e. the highest level of inference). For visualization purposes, no differentiation is made between predictive and prediction error nodes or clusters (for details on this, see Fig. 6). Individual nodes in this figure may represent both single nodes and clusters, conforming to the scale invariant principle.

In hierarchical Bayesian inference, each superordinate level performs a form of 'pattern recognition' on events that take place at subordinate levels. The superordinate level thus encodes a more generalized and parsimonious (abstract) model of events that happen below it. Such higher-level generative models go well beyond the lower-level data that helped to spark their existence: they may involve quite creative designs that may autonomously inform behavior (Tenenbaum et al., 2011). When this principle is applied systematically to goal states, something interesting happens. As mentioned, organisms produce a hierarchy of goal states that eventually involves a global division between internal (self-referential) and external (social) goal states, both of which can be set proximally or more distally in time. Logically then, the hierarchical integration across goal states can be pushed one level further, involving an additional level of inference across these two global goal states. This produces an overarching third series of goal states that are common to both the organism itself and its (social) environment, across timescales (Fig. 7).

Such models transcend the level of the individual organism, its immediate (social) environment, as well as the immediate moment. In other words, such goal states define (social) laws, rules or standards that hold across different individuals, social groups and timescales (Constant et al., 2019; Toelch and Dolan, 2015). Thus, hierarchical Bayesian inference predicts that, eventually, organisms produce goal states that they consider to have general validity for everyone across (infinite) time. Prediction errors that are produced relative to such 'normative' goal states may involve a time and energy investment that is not immediately contingent to the interests of the organism itself. Rather, such behavior is aimed at striking a balance between the short-term and long-term interests of individuals and ever more distant social groups (including future generations), i.e. to promote global rather than local stability. Individuals that follow such goals will at times make decisions that favor the (long-term) interests of others rather than themselves, i.e. they will show altruistic behavior. Additionally, such goals may cause some members of a group to punish themselves or others for social norm violation (Fehr and Schurtenberger, 2018). Altruistic and law-abiding forms of behavior have been observed in a variety of (higher) organisms (e.g. Bekoff and Pierce, 2009). We expect such goal states to represent the highest level of hierarchical Bayesian inference and, therefore, the highest level of integration within a nested hierarchy of network clusters. In other words, they truly represent our 'highest goals'.

This prediction can be tested by examining organisms that differ in the degree to which they engage in activities that are aimed at promoting global and long-term rather than local and short-term stability of individuals and groups (e.g. mediation versus social polarization, fairness versus unfairness in the sharing of energy and resources, punishment for social norm violation versus laxity, altruistic versus selfish behavior, transpersonal identification versus nepotism, transgenerational identification versus generational individualism, etcetera). Combinations of such functions are typically (but not exclusively) found in so called 'higher organisms', and should link to measures of hierarchical depth in nested modular biological networks (see Discussion). Thus, hierarchical Bayesian inference may explain why higher organisms tend to have bigger throughput areas (e.g. the giant nucleus of eukaryotic versus prokaryotic cells, or the frontal and anterior extensions of the brains of higher primates): such hierarchies are required to accommodate more encompassing world models. Despite such extensions, however, the basic principles that govern behavior in higher organisms appear to be the same as in woodlice: action sequences are produced that aim to minimize prediction error relative to world models with different degrees of model complexity (Botvinick and Weinstein, 2014; Friston, 2012).

In humans, empirical studies of goal states have produced a hierarchical taxonomy that eventually involves the global goals of agency (connecting with the self), communion (to connect with a local social group) and meaning (connecting across spatial, temporal and social barriers; Talevich et al., 2017). These global goals are closely related to Maslow's hierarchy of needs (with multiple levels of self-actualization, social belonging and transcendence) (Koltko-Rivera, 2006). Such goal states have a strong resemblance to internal (self), external (social) and cross-cutting (normative) goal states as predicted by hierarchical Bayesian inference. By now, the human mental phenotype has been mapped quite well with respect to the presence of normative functions and individual differences in the degree to which subjects score on these phenotypical dimensions can explain differences in normative or moral behavior (e.g. Koltko-Rivera (2006); Stankov (2007); von Collani and Grumm (2009)) as well as individual differences in brain structure and function (see below). The existence of such domains of functioning has been eschewed by scientists for quite some time because of its inherently moral (or even religious) nature. Nevertheless, such domains are predicted by the principle of hierarchical Bayesian inference and supported by evidence from various domains of science.

In the active inference literature, goals are prior beliefs about controllable factors in the environment that rest upon each organism's place in a particular eco-niche, with each niche showing varying degrees of pro-sociality. This leads to the notion of variational eco-niche construction, whereby each individual builds its own generative models that can be shared among other members of its family or conspecifics (Constant et al., 2018; Veissière et al., 2019). The notion of higher (interpersonal) goals amounts to a shared generative model or narrative that ensures the members of a group can predict each other - and thereby minimize their prediction errors. In humans, the need to exchange such higher-level insights gives rise to our scientific, moral and legal institutions, which may aid in the attempt to eventually construct a globally held world view that serves to optimally inform human behavior. Below, we will discuss how stress causes organisms to downgrade on contextual processing (model complexity) and discuss its impact on (human) behavior.

### 3.5. Stress in hierarchical Bayesian control systems

The hierarchical Bayesian control systems perspective on living organisms allows for a clear definition of 'stress' (Peters et al., 2017). Stress can be defined as the difference between some desired or anticipated state (a setpoint, goal state, or world model) and the actual input state of an organism. Mathematically, this can be framed as the differences between empirical priors and posterior expectations, i.e. the overall level of prediction error. Likewise, the term *response* can be defined as the behavior that follows these prediction errors (note that according to this definition, every prediction error is a form of stress, and any response can be defined as a stress response). (Stress) responses serve to counter the perturbation of a control system and allow the system to return to a more stable state. Since organisms occupy a wide range of environmental niches in which they face a multitude of unique stressors (e.g. specific chemical constitutions, rivals or predators), stress responses are in many cases unique and involve unique messaging pathways. Some stress responses are more general, however, such as the stringent response in bacteria when subjected to nutrient deprivation (Boutte and Crosson, 2013), or the SOS response in case of DNA damage (Baharoglu and Mazel, 2014). Such 'general stress responses' are mounted in a similar way across species, regardless of the specific stressor the organism encounters (e.g. starvation, drought, heat, cold, acidity, salinity, DNA damage, social stress) and involve the up- or downregulation of a few key transcription factors that have been tightly conserved throughout evolution (de Nadal et al., 2011d; López-Maury et al., 2008; Lyon, 2015; Marles-Wright et al., 2008; Nagar et al., 2016; Storz and Hengge, 2010). Whereas the upstream changes in transcription factors that occur during general stress responses show a clear overlap between species, however, the downstream changes appear to be more species-specific. Additionally, general stress responses have only been characterized in a small number of model organisms, so it is uncertain whether they occur within all species of bacteria, or indeed in other species (Gottesman, 2019). At first glance, then, the sheer heterogeneity of stress responses seems to deny the existence of a truly 'general stress response'. Despite such heterogeneity, however, recent studies found evidence that some aspects of the stress response are indeed universal across species, whether they be bacteria or plants, mammals or humans. This condition- and species agnostic response can be quantified in terms of the overall amount of regulatory activity that takes place under stressful conditions. Studies show that bacteria that are challenged with an evolutionary familiar stressor show subtle responses of gene transcription, whereas bacteria that are challenged with a relatively unfamiliar stressor (e.g. an antibiotic) show larger and seemingly more chaotic responses (Jensen et al., 2017). More specifically, stressed bacteria express a larger number of different genes with increasing amplitudes, while gene expression is becoming increasingly uncoordinated as the challenge endures. These changes have recently been quantified in terms of entropy, which is a well-established information theoretic quantity of disorder (i.e. $H = \ln(|M|)$ where M is the (permuted) covariance matrix containing gene co-expression strengths) (Zhu et al., 2020). Rising entropy levels in the signaling pathways of bacteria successfully predict bacterial fitness in terms of growth rate (stagnation) and survival (death) under stressful conditions. This is true regardless of the specific environmental conditions, the types of genes that are involved and the strain or species of bacterium under study. Rising entropy levels have been used to predict the success of antibiotic therapy for any type of antibiotic in any strain of bacterium, which is far more efficient than current gene panels that rely on specific gene expression profiles in specific micro-organisms (Zhu et al., 2020). The increased amount of disorder in gene expression profiles that is observed in stressed bacteria has been explained in terms of a loss of regulatory influence ('dysregulation'), which normally coordinates dependencies between genes and produce some degree of order. The loss of such coordination then causes gene expression levels to vary independently and more randomly. In other words, rising levels of entropy seem to signal a regulatory overload, which is predictive of a loss of fitness.

The predictive power of (permutation) entropy or similar measures generalizes well beyond bacteria. It has been used to predict behavioral changes of a large number of different classes of organisms under stressful conditions, including plant species (Sun et al., 2010), fish and other aquatic organisms (Bae and Park, 2014; Eguiraun et al., 2014), insects (e.g. Liu et al., 2011), chickens (Marıa et al., 2004,), quails, rats, pigs and primates, to name but a few (e.g. Asher et al., 2009). Interestingly, increased disorder has been discovered in timeseries of human behavior under stressful conditions. Human inner experience and overt behavior (the mental phenotype) can be measured using experience sampling methodology (ESM): a technique that involves rating multiple phenotypical items several times a day for several weeks or months to produce timeseries. When stress levels increase, typical changes can be observed in such timeseries that involve increased levels of variance, increased amplitudes, increased anticorrelations between opposing mental states (e.g. happiness and sadness), increased temporal auto-corrrelations and a slow recovery from external perturbations (van de Leemput et al., 2014). Together, these changes signal the phenomenon of 'critical slowing down' (CSD), which is a highly generic state of network systems that are poised on the brink of a 'tipping point'(a sudden transition from one state of the system to another). Just before the onset of such phase transitions, the systems starts to show erratic behavior (CSD). CSD is a generic characteristic that can be used as an early warning sign to predict the occurrence of tipping points in non-living as well as in living systems (Veraart et al., 2012; Scheffer et al., 2012). In humans, CSD has been used successfully to predict the onset of a mental disorder (major depression) at least 3 months in advance (van de Leemput et al., 2014). As is evident from their respective definitions, CSD is actually synonymous with a (transition towards a) state of high (permutation) entropy. Entropy is a much more general term, however, which can be quantified from timeseries data using a single parameter instead of three terms or more ($H = \ln |M_p|$, where $|M_p|$ denotes the determinant of a graphical lasso regularized empirical correlation matrix), or even from a single timepoint (i.e. $H_{stp} = \ln(\sigma^2)$, where $\sigma$ refers to the measured variance in the expression of recorded variables for a single timepoint) (Zhu et al., 2020). In short, a universal stress response can be formulated not by looking at the specifics of regulatory activity in living systems, but rather at the total amount of disorder observed in hierarchical message passing within organisms (as measured e.g. by a bacterial transcriptome or brain activity). Rising levels of entropy have been proposed to result from a loss of 'regulatory connections', which normally coordinate (e.g. synchronize) the different elements of the system and produce order (Zhu et al., 2020). High levels of permutation entropy can serve as a generic early warning sign for sudden state transitions reflecting a failure of control, which signal either stagnant growth, disease or the death of an organism (i.e. a loss of homeostasis). We will now examine whether the overt behavior of organisms under high levels of stress shows universal changes as well, in order to derive a general theory of stress and the stress response in living control systems.

When studying the overt behavior of organisms under high levels of prolonged stress, features emerge that appear universal to all organisms. Whereas short or sublethal stress levels seem to speed up metabolism, promote motility (fight or flight) and enhance exploration tendencies (migration), social activity (establishing hierarchy), the exchange of genetic material (procreation) and parental investment in a wide range of organisms, prolonged and (near) lethal stress levels induce behavioral changes that involve a down-regulation of metabolism (e.g. bacterial stasis, sporulation, hibernation), reduced motility (mobility or migration), reduced sociability, a halt on reproductive activity, an increase in (DNA) repair activity or sleep, and a tendency to neglect (abandon, or even eat) offspring (Wingfield et al., 1998, 2003; Ruf and Geiser, 2015; Hausfater and Hrdy, 2017; Del Giudice, 2020). Such 'emergency life history stage responses' generally economize on long-term, (pro)social and/or reproductive activities in favor of short-term, self-repairing and self-preserving activities. In more consize terms, severe stress is said to cause organisms to shift away from 'slow' policies (i.e. long-term prosocial activities and parental investment) and towards 'fast' policies (i.e. short-term and self-preserving activities) (Del Giudice, 2020). Such shifts in behavioral policies are especially evident in social species (when compared to solitary species), since these normally devote a significant amount of their time in building social hierarchies and parental investment (Del Giudice et al., 2015). Nevertheless, even bacteria are known to cut down on 'social' and reproductive activities in response to a (near) lethal stressor, e.g. when shutting down horizontal gene transfer, halting cell division or engaging in sporulation (Lyon, 2015; Meeske et al., 2016). It therefore seems that organisms upregulate complex behavioral policies under intermediate levels of stress but abandon such policies when stress levels approach near lethal levels. Such behavioral changes have been explained in terms of 'allostatic overload', which refers to the situation where the regulatory capacity of a control system is overtaxed by environmental perturbations, i.e. where regulatory work increases to the point where energy demand exceeds energy supply (Wingfield et al., 1998, 2003, McEwen and Wingfield, 2003). In such cases, organisms need to cut down on computationally expensive regulatory activities in order to save energy and resources. Interestingly, scholars have linked the expenditure of energy and resources to the hierarchical depth of information processing (Hermans et al., 2014; Goelzer and Fromion, 2017; McEwen and Wingfield, 2003): higher ('allostatic') levels of hierarchical control systems that inspire more complex forms of behavior demand more energy, whereas lower ('homeostatic') levels that control relatively simple behavior require less regulatory work and demand less energy (see introduction). Thus, organisms appear to abandon hierarchically higher levels of processing in favor of lower processing levels when confronted with allostatic (higher regulatory) overload (Hermans et al., 2014). This has been demonstrated experimentally in different organisms including humans, e.g. with human behavior falling back from goal-directed to habitual behavior under severe levels of stress (Schwabe and Wolf, 2011; Goelzer and Fromion, 2017; Van Oort et al., 2017). In the previous section, we saw that self models, social models and transcendent (normative) models involve the highest levels of contextual integration (in a spatial/social and temporal sense) and inspire complex forms of goal-directed behavior, including moral decision making. Consistent with the theory of allostatic overload, severe stress is known to negatively affect prosocial behavior and moral decision making (although moderate levels of stress may actually increase prosocial behavior, see below) (Lee and Yun, 2019; Mendez, 2009; Starcke et al., 2011; Youssef et al., 2012). Extreme stress therefore seems to affect policy selection as a function of contextual integration, i.e. organisms take lesser amounts of contextual information into account when formulating a response. Such decontextualization allows them to revert to more basic policies that demand less energy. In short, increased levels of entropy in hierarchical message passing within living systems are already suggestive of a loss of integrative control under severe levels of stress, but this notion is further backed by studies of the overt behavior of stressed organisms that independently point towards a reduction in model complexity and corresponding shifts in policy selection. We therefore propose that extreme stress causes a top-down collapse of goal hierarchies, i.e. a loss of hierarchical depth. This forces organisms to downgrade from high-level (functionally integrated) goal states to lower-level (functionally segregated) goals and corresponding behavioral policies. The universal presence of this principle suggests it has cornerstone value in securing survival 'when the going gets tough'.

For a mechanistic account on how this might work, it is worthwhile to examine the biophysics of stress in hierarchical (Bayesian) control systems. In such systems, prediction errors are used in two distinct ways (Fig. 6): to update Bayesian beliefs by resetting priors (changing your mind) and to induce an output sequence or stress response to reduce the error via the environment (changing your actions). Lower level policies (e.g. walking) are allowed to run freely until prediction errors are produced within the input hierarchy (e.g. stumbling across some unforeseen object; Scafetta et al., 2009). If error signals cannot be sufficiently suppressed by a simple, straightforward response generated at some lower level of the hierarchy (e.g. side- stepping), the residual error is 'escalated upward' into the hierarchy to update more comprehensive world models and produce a corresponding, more complex output (e.g. walking around the object; de Kleijn et al., 2014d). Thus, prediction errors pass a hierarchical succession of goal states (increasingly complex generative models) and corresponding behavioral strategies until they are suppressed. Such elaborate strategies may eventually suppress prediction errors in ways that simpler forms of behavior cannot (e.g. by successfully walking across the object, reaching the top of a fruit tree, or climbing a social hierarchy). This may explain why intermediate levels of stress initially cause organisms to display more complex behavioral strategies. The vertical accumulation of prediction error can be thought of in terms of a loss of control over free energy. In the Bayesian inference literature, rising levels of free energy are usually associated with increases in entropy and a concomitant loss of thermodynamic and computational efficiency. The use of higher order (more complex) strategies is therefore likely to coincide with rising levels of (permutation) entropy in measures of hierarchical message passing and overt behavior.

Since any hierarchy of control systems is finite, however, prediction error signals may accumulate upwards across multiple levels of control until the top of the hierarchy is reached. At that point, the organism has exhausted its hierarchy of goal states and corresponding policies (i.e. even complex strategies are ineffective at suppressing prediction errors). In such cases, vertically accumulated prediction errors activate a small number of hub structures located near the top of the (goal) hierarchy (the knot of the bow-tie). These hubs maintain many long-distance connections with other network clusters and subclusters in the network, thus representing the highest level of integration within the network structure at large (Figs. 5–7). From simulation studies in statistical physics, it is known that the highest degree nodes in a network have the highest levels of energy dissipation, corresponding to highest energy demand (Gosak et al., 2015). The fruitless pursuit of high-level goal states and corresponding behavioral policies may therefore cause these structures to be flooded with ascending prediction error signals, to the point where energy demand exceeds energy supply. When this happens, these high-level hub structures will overload and fail (Gosak et al., 2015; Stam, 2014). This 'allostatic overload' has been experimentally confirmed to coincide with increased levels of permutation entropy specifically for such hub nodes (Sun et al., 2010). Since high-level hub structures normally integrate information streams across a large number of subordinate clusters (functional integration), their shutdown causes a shift in the balance between functional integration and segregation of network clusters in favor of functional segregation (Tononi et al., 1994). This corresponds to a collapse of the nested modular goal hierarchy: more encompassing goal states effectively 'decompose' into their constituent components, inducing a corresponding change in behavior. As a result, the subordinate network clusters will no longer be functionally connected and start to operate independently (functional segregation). This loss of integrative control and the ensuing uncoordinated activity of subordinate modules will add significantly to increased scores on (permutation) entropy.

To be clear, we wish to emphasize that hub overload and failure is most likely to involve a *functional* shutdown of high-level predictive hub structures and decreased functional connectivity, rather than a structural loss of hubs and a loss of *structural* connectivity (i.e. failing hub structures still remain physically in place). At least, this seems to be the case in acute forms of stress. In chronic forms of stress, studies show that even a loss of structural connectivity may occur (e.g. synaptic pruning), which may involve the active degradation of maladaptive world models (goal states and corresponding policies) by means of substances such as glucocorticoids (Peters et al., 2017). We predict that the collapse of goal hierarchies is a function of node degree: the most integrative goal states are the first to go, but subordinate levels with lesser-degree hubs and corresponding subgoals may follow depending on the amount of accumulated stress (prediction error). Severe stress may therefore cause a graded disintegration of a nested hierarchy of goal states across several levels. Like military command collapsing in a top-down fashion (generals first, then colonels, lieutenants, higher officers, etcetera), allostatic overload may dissolve goal hierarchies, leaving only the local troops and the odd sergeant major to take care of the problem (Fig. 8). This may explain why severe levels of stress eventually cause organisms to display increasingly primitive forms of behavior. This hypothesis can be tested by examining measures of hierarchical depth of (functional) network structures in relation to policy selection and behavioral complexity at different levels of stress (see Discussion).

When higher-level hub structures overload and fail, they lose their influence as empirical priors that are important in maintaining the balance between top-down prior beliefs and bottom-up sensory evidence (Fig. 6). The overall *amplitude* of prior signals or prediction errors is often quantified in terms of 'precision', which refers to the inverse variability (dispersion) of a probability distribution. In other words,

A.

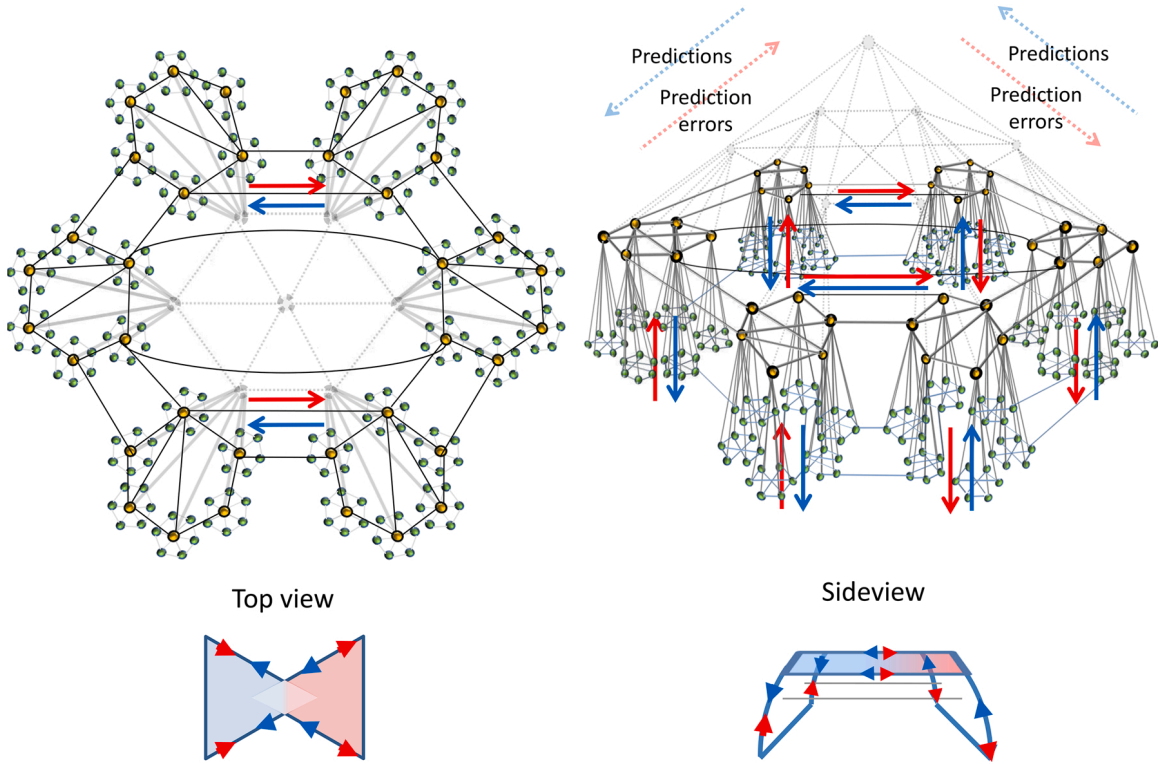

Top view

Sideview

B.

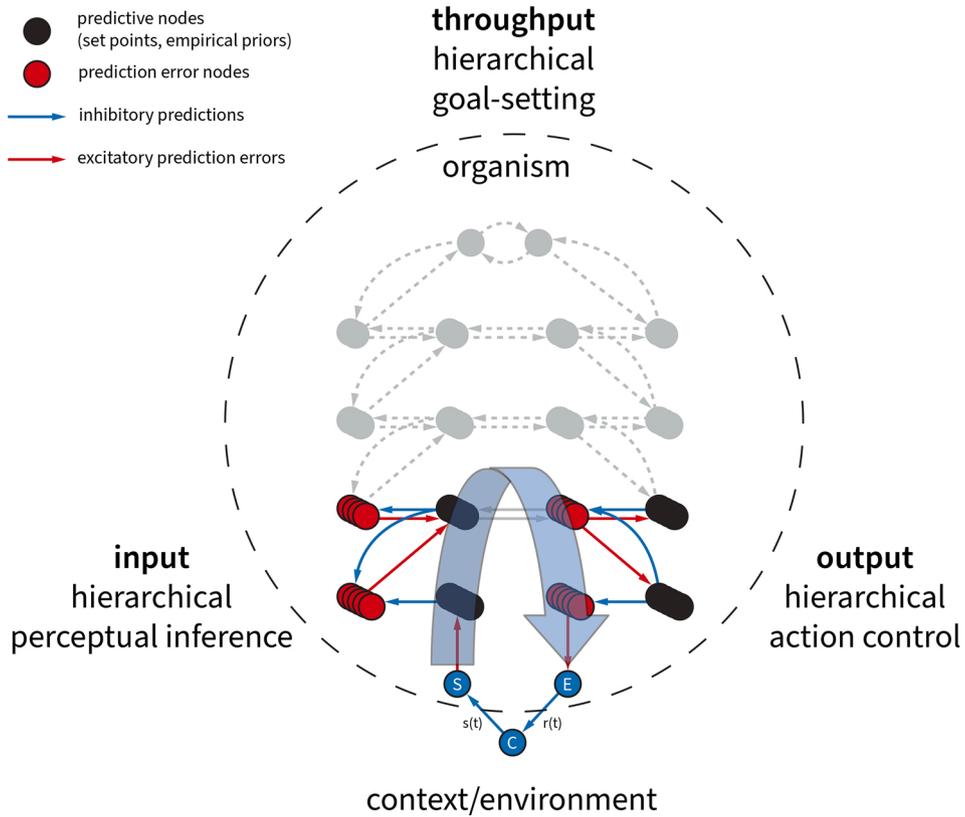

*(caption on next page)*

**Fig. 8.** Severe Stress in Organisms: The Collapse of a Hierarchy of Goal States.

*Note:* Hierarchical Bayesian control systems allow organisms to incorporate an increasing number of contextual cues from their environment and create a hierarchy of 'world models', i.e. goal states that are used to inform behavior. At the highest levels of integration, such goal hierarchies involve internal (self) models, external (social) models and cross-cutting (normative) models (Fig. 7). In severely stressed organisms, this goal hierarchy collapses in a top-down manner, possibly as a result of hub overload and failure (grey nodes). This results in a 'decontextualization' of behavior, with organisms favoring short-term and self-centered policies (informed by self models) over long-term social and/or normative behavior (social and normative models), to save energy and resources. The regulatory collapse may involve several hierarchical levels of integration, depending on the error levels that are encountered. Phenotypically, this manifests as organisms 'downgrading' from goal-directed to instrumental, habitual or even reflexive forms of behavior. The top-down loss of hierarchical control by high-level (inhibitory) empirical priors produces both a disinhibition and loss of coordination of lower levels, adding to permutation entropy levels. Please note that individual differences may cause some organisms to retain top-down control under severe levels of stress. See text for further details.

precision scores the 'confidence' in a Bayesian belief or prediction error. Increasing the gain of prediction units means that these priors are selected, causing them to more strongly suppress prediction errors and render them more resistant to belief updating. Conversely, increasing the connective efficacy (amplitude or 'gain') of prediction error units means that the associated prediction errors are selected, enabling them to preferentially induce belief updating higher in the hierarchy. This leads to the notion of precision at different levels of the hierarchy, whose balance is crucial for determining the relative influence of top-down prior beliefs relative to bottom-up sensory evidence. We can therefore think of stress as reducing **prior** precision and rendering the organism more exposed to belief updating based upon immediate sensory evidence: stress alters the connective efficacy of priors and prediction error units, thereby sequestering them from other levels of the hierarchy. This collapse can either be reversible, e.g. reflecting modulatory control of connection strengths in acutely stressful situations. Alternatively, it could be mediated by long-term changes in connective efficacy or a loss of connections per se, which have been reported in neural systems after chronic stress[3] (e.g. McEwen et al., 2015).

In short, stress seems to change behavioral policies according to a hierarchical principle, i.e. increasingly less contextual cues are used to inform behavior, suggesting a top-down collapse of goal hierarchies. This 'decontextualization of behavior' has several short-term advantages. First, organisms will spend less energy and resources on reaching long-term and complicated goals, which allows them to endure current unfavorable conditions for longer periods of time. Second, the bypassing of higher-level systems reduces the path length of the network, allowing signals to travel from input to output structures across shorter distances, producing faster responses (computationally, this is equivalent to minimizing model complexity or computational complexity costs). Third, the top-down collapse of integrative control reduces the gain of predictive connections that normally constrain (inhibit) lower-level policies. This allows such policies to be expressed more freely, making them more pronounced and easy to trigger. This is referred to as 'disinhibition' in the psychological sciences and involves a heightening of the senses (within the input hierarchy) and a strengthening of responses (within the output hierarchy), to produce a 'livening of the reflexes' (Gorenstein and Newman, 1980). Thus, organisms capitalize on model complexity and precision to formulate a stronger and faster response. This may provide organisms with just the edge needed to force themselves a way out of a dire situation (Byrd et al., 2019).

Such changes come at a price, however, which is a loss of regulatory finesse. A reduction of model complexity makes organisms more vulnerable to environmental conditions that require a broader (and/or long-term) perspective. Additionally, a deep collapse of a regulatory hierarchy may lead to a state of disinhibition where any input almost immediately triggers a strong output and vice versa. In such a case, even a small environmental disturbance may trigger an intense, volatile, and uncoordinated response (Byrd et al., 2019). This response may then change the environment of the system to the effect that it serves as a trigger for a novel response, and so on. The self-sustaining (circularly causal) pattern of reflexive activity that thus emerges is called a 'clonus'. This refers to situations where a loss of higher-order inhibitory constraint causes the input and output elements of a control system to become strongly coupled (i.e. the intrinsic coupling of the system is enhanced, causing it to become strongly reactive to input). Such a strong intrinsic coupling then induces a stronger extrinsic coupling (of the organism with its environment) and clonic activity. At some point, the system may become so strongly coupled to its environment that it will lose its ability to compensate for environmental disturbances: it will decompensate (lose control), after which the interior state of the system will linearly follow that of the environment (i.e. a loss of homeostasis). In living systems, such tipping points amounts either to disease, or the death of the organism.

To our knowledge, this is the first detailed model of allostatic overload, or the way in which stress may cause a top-down collapse of high-level integrative control that leads to increased levels of disorder (entropy) in hierarchical message passing and overt behavior in living systems, to eventually produce tipping points (disease or death). Such tipping points occur when such a collapse reaches too deeply down a hierarchy of control systems (i.e. when a hierarchical tree is pruned beyond a level of adequate control). According to this model, organisms may differ in their susceptibility to tipping points as a result of individual differences in the outgrowth (maturation) of their regulatory hierarchies, i.e. different heights of the regulatory tree come with different thresholds for tipping points (decompensation) and, hence, biological fitness. This hypothesis can be tested e.g. by examining the degree to which measures of the hierarchical depth of biological networks predict entropy levels and tipping points under varying levels of stress (see Discussion).

Previously, scholars have defined stress specifically as a failure of control (e.g. Del Giudice et al., 2018), but provided no clear mechanism. Others focused more on physiological states (McEwen and Wingfield, 2003) or cognitive processes in humans (Koolhaas et al., 2011; Ursin and Eriksen, 2010). Most previous definitions of stress situate that state somewhere in between criticality and tipping points as defined above. Here, we employed a more liberal definition of stress as the (cumulative) error state of hierarchical control systems (Peters et al., 2017). The advantage of this definition is that is can be generalized across species and that it lies on a continuum, with clear and objectifiable stress-responses marking discrete levels of stress, i.e. (0) Routine performance (low levels of prediction error, low entropy, reflexive, instinctive (Pavlovian) or habitual behavior, 'homeostatic control'), (1) Creative problem solving (upward escalation of prediction error signals, rising entropy, goal-directed action, 'allostatic control'), (2) Emergency responses (high levels of prediction error, high entropy, top-down collapse of goal hierarchies, 'allostatic overload', downgrading from goal-directed to lower forms of associative learning, 'regression to homeostatic control'), (3) Critical slowing down (high prediction error, high entropy, near loss of control) (4) Tipping points (decompensation, loss of control). For a similar categorization of the stress response, see

---

[3] Neurophysiologically, precision is usually thought to be mediated by the control of synaptic efficacy; either through neuromodulatory transmitter systems or the nonlinear dynamics that mediate synchronous gain. This will be particularly relevant later when we talk about psychopathology. This follows because most of the drugs used in psychiatry act upon the neurotransmitters that modulate synaptic gain and therefore control the precision of message passing in the human brain.

(Romero et al., 2009).

## 4. The human brain as a hierarchical Bayesian control system

So far, we have discussed rules of network structure and function that may apply to all living systems. We will now show that such rules apply to human behavior. At larger spatial scales, the human brain has a multimodular, hierarchically controlled *small world* network structure (Bullmore and Sporns, 2009; van den Heuvel et al., 2008v). Its 86 billion neurons (Azevedo et al., 2009) form neural modules that are an average of around 5 degrees of separation apart from any other module in the brain (Bassett and Bullmore, 2006; Hilgetag and Goulas, 2016; Sporns and Zwi, 2004; van den Heuvel et al., 2008v). These modules form a nested hierarchy of part-whole relationships (Meunier et al., 2010, 2009) that give rise to a bow-tie network architecture (Markov et al., 2013). Perceptive areas form the input hierarchy of this bow tie, the medial (pre)frontal lobe and anterior insula its knot and (pre)motor cortices and hypothalamic areas make up the output hierarchy. Overt behavior reflects the concerted action of large numbers of simple input-output patterns at the bottom of this hierarchy ('reflexes', which tie basic input to motor and endocrine output primitives), the activity of which is carefully orchestrated by higher levels of integration (Fig. 8; Botvinick, 2008; Botvinick and Weinstein, 2014; Freeman, 2001, 2005; Ribas-Fernandes et al., 2011). The human brain has been compared to a Bayesian inference engine, whose primary job it is to infer the (hidden) causes of its sensory input by building predictive models of the world and acting upon those models (Friston, 2010; Friston et al., 2006). In doing so, a generative model is constructed with multiple hierarchical levels of model complexity that constitutes our inner experience and overt behavior (the human mental phenotype, or 'mind'). Fig. 9 summarizes current ideas on the human mind as a hierarchical generative model that has its origin in different forms of perceptive information (Badcock et al., 2019; see below for further references). The statistical structure of this phenotypical hierarchy is assumed to mirror that of the human brain (i.e. it has a nested modular bow-tie network structure).

At the bottom of the phenotypical hierarchy, three global types of perceptual input can be discerned. Exteroceptive perception involves information coming from the external environment, i.e. the main senses

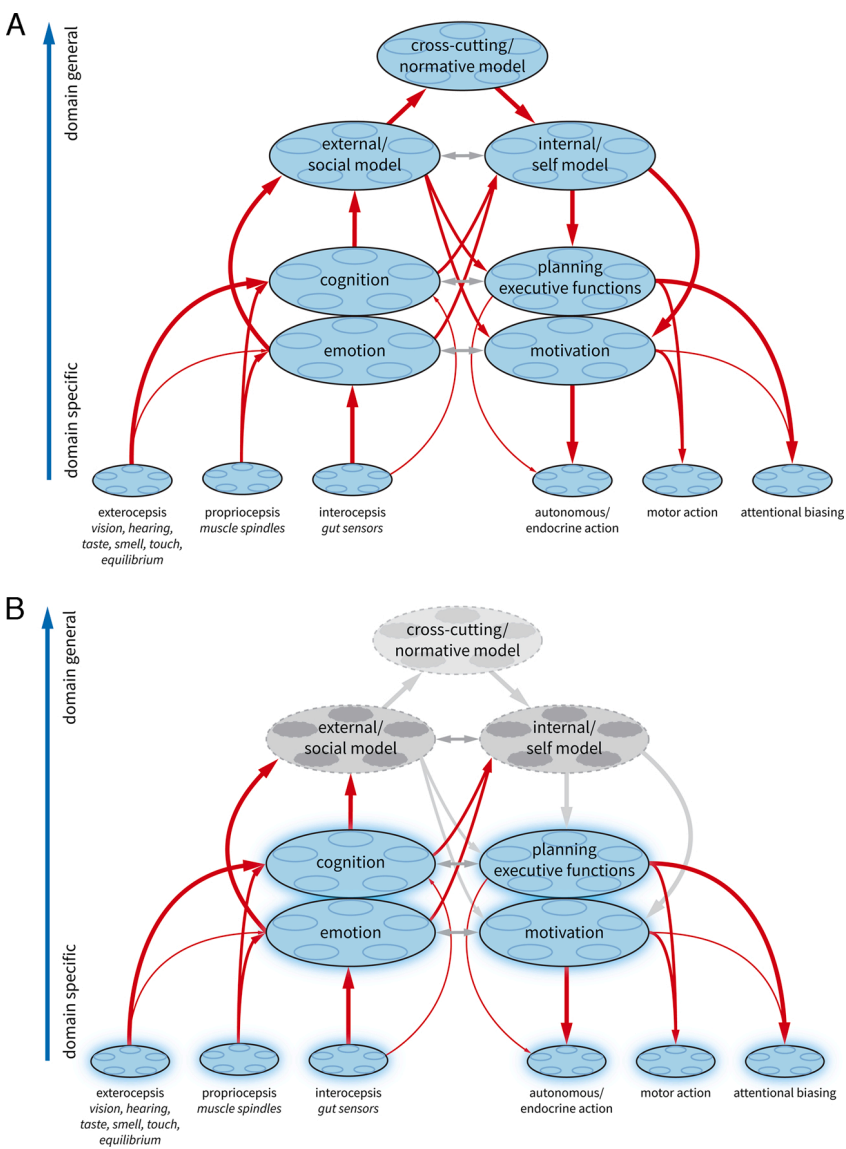

**Fig. 9.** The Human Mind as a Hierarchical Generative (Bayesian) Model.

*Note:* **A.**The human mind can be modeled as a nested modular hierarchical generative model that controls perception and action (Badcock et al., 2019). This figure summarizes current ideas on the statistical dependencies between the different components of this hierarchy, which are assumed to echo those of human brain (i.e. a nested modular folded bow-tie structure). Circles indicate generative models and circles within circles subordinate models. Higher level (domain general) models are inferred from progressively lower (domain specific) models that eventually have their origin in different forms of perceptive information (e.g. exteroceptive, proprioceptive and interoceptive domains). Arrows sizes reflect the putative contribution of a particular domain in biasing inference within another domain (see text for further details). Note that cognition, emotion, executive functions and motivation occupy a similar hierarchical level of inference (the Figure is 3D). Domains may affect each other across loops of different pathlengths (e.g. from input to output via a hierarchically ordered set of goal states), reflecting different degrees of information processing (policy selection). The shortest loops within this hierarchy represent basic stimulus-response patterns (e.g. simple and more complex 'reflexes', Pavlovian instinct patterns and habitual behavior), whereas the longest loops reflect goal-directed behavior that is informed by highly integrated world models involving self-referential, social and normative models (Fig. 6). Each phenotypical domain may have multiple functional-anatomical brain regions as a correlate (see text and references for further details). **B.**When stressed severely, contextually redundant higher-level goal states are shut down to save energy and to enhance the stress response (Fig. 8). This corresponds to a collapse of self, social and/or normative models, causing a shift away from goal-directed behavior (longer loops) towards habitual, instinctive or reflexive behavior (progressively shorter loops). The loss of higher-level integrative constraint triggers a disinhibited and disordered state at lower levels within the hierarchy (glow), involving emotional, motivational, cognitive, perceptive, premotor executive and action domains. A shallow collapse may provide leverage out of a difficult situation, but a deep collapse will cause the current model to revert to a hierarchical model of psychopathology (HiToP, see text). In such cases, the system will show increased intrinsic connectivity, which enhances the extrinsic connectivity of the individual, i.e. an increased dependence on the environment and decreased homeostasis. This may present e.g. as strong interpersonal dependencies and/or social conflict. When goal hierarchies fail to mature, such underregulated states become chronic (e.g. personality disorders). See text for details.

of vision, hearing, touch, smell and taste. Interoceptive information involves information feeds coming from the internal environment, e.g. gut and vascular pain and blood pressure afferents, blood glucose concentrations, smooth muscle tension, et cetera. Finally, proprioceptive information takes up position in between the internal and external environment and mostly involves input from skeletal (striated) muscles, tendons and bones. In is thought that hierarchical inference on these basic input domains progressively produces the human mind (Badcock et al., 2019, Fig. 9). Recent studies conceptualize human emotion as hierarchical Bayesian inference on predominantly interoceptive information, placing this hierarchy of affective generative models somewhere along the middle of the larger hierarchy (Seth and Friston, 2016; Smith, 2019a). Similarly, cognition may involve hierarchical inference on predominantly exteroceptive information (Smith et al., 2019b). Executive functions in turn involve part of an output hierarchy that is engaged in high-level (conceptual and posterior) planning, with a predominant connection to motor output (controlling muscle action) (Pezzulo, 2012). Motivational functions have been conceptualized as aiding in predicting the precision of motor and endocrine output, with a possible emphasis on endocrine action (Pezzulo et al., 2018). As discussed above, the top of this hierarchy involves highly integrated generative models of their inner (self) and outer (other) states, along with their histories and possible futures. Self models are processed in midline areas of the human brain, which are involved in some of the highest levels of integrative processing (Haggard, 2017; Northoff et al., 2006; van der Meer et al., 2010v; Thornton et al., 2019). Additionally, humans make highly integrated models of the states of others and their possible histories and futures. Such social models (or 'theories of mind') involve medioprefontal, (superior) temporal and temporoparietal areas (Amodio and Frith, 2006; Gallagher and Frith, 2003; Mars et al., 2013; Thornton et al., 2019), which process information at very high levels of contextual integration. Finally, a large body of literature has identified brain regions that are involved in making decisions about events that go beyond the self or the immediate social environment, but instead involve common social laws and values. These normative structures include ventromedial areas for norm processing and right insula, dorsolateral prefrontal, and dorsal cingulate cortices for processing in relation to social norm violation (Zinchenko and Arsalidou, 2018). Such brain areas again involve some of the highest levels of integration across subordinate systems. Together, such studies provide both phenotypical and neuroanatomical support for the existence of a hierarchy of generative models with interior (self), exterior (social) and normative structures at the top of this hierarchy.

Overall, our brains seem to have capitalized particularly on information processing at high levels of functional integration, making detailed predictions of events that take place more distally in time as well as in (interpersonal) space (Herrmann et al., 2007). The ability of the human brain to take large amounts of contextual information into account when formulating a response seems to explain much of its disproportionate size (Dunbar and Shultz, 2007). Despite such extensions, however, the basic principles of control theory that govern behavior in lower organisms remain the same as in humans. As in woodlice, activity levels drop (i.e. we become quiet and pleased) when the perception of our past, current and future environment agrees with our intricate interpersonal goals and expectations.

When observing human brain function and behavior under severe levels of stress, several things stand out. Although mild forms of stress differentially affect or even enhance our personal sense of identity, promote social cohesion or a sense of global connectedness, severe stress brings us into 'survival mode' (Buchanan and Preston, 2014; Mao et al., 2016; McEwen and Wingfield, 2003; Von Dawans et al., 2012). Neuroimaging studies show that the human brain falls back from goal-directed to habitual control during severe stress (Schwabe and Wolf, 2009, 2011). This corresponds to decreased activity in higher level systems such as the anterior cingulate, anterial insular and temporopolar areas (Arnsten, 2009; Dias-Ferreira et al., 2009; McEwen et al., 2015;

McTeague et al., 2016; Schwabe and Wolf, 2009, 2011; Van Oort et al., 2017). Brain areas that decrease activity during severe stress are midline structures involved in generating self-models (self-image; Goette et al., 2015; Hooley et al., 2013; Kesting et al., 2013; Staniloiu and Markowitsch, 2012), as well as brain areas associated with the production of social world models or theory of mind (Sandi and Haller, 2015; Todd et al., 2015), producing more selfish forms of behavior. Finally, severe stress is known to negatively affect moral decision making (Lee and Yun, 2019; Mendez, 2009; Starcke et al., 2011; Youssef et al., 2012). This change in behavior is related to altered activity in brain areas involved in transpersonal identification, including law-abiding and moral behavior (Lee and Yun, 2019). Thus, severe stress decreases activity specifically within brain areas that support some of the highest forms of contextual integration. Such findings support the hypothesis that significant stress causes a top-down collapse of deep goal hierarchies to save energy and resources, causing people to take increasingly less amounts of contextual information into account when formulating a response (Figs. 8 and 9). Of course, individual differences may cause some people to deviate from this general pattern.

In short, we propose that severe stress prunes the top of a regulatory pyramid in people's brains to produce a subtle form of decortication (hypofrontality, or a 'chicken without a head' syndrome). Such a top-down collapse of goal hierarchies reduces the amount of integrative control (lowers the gain of inhibitive empirical priors), which increases disorder at subordinate levels of the hierarchy, down to the level of the shortest reflex loops. This may manifest as a more violent expression of behavioral primitives or stress response patterns such as fight, flight, fright, feeding, freezing, reproducing, fainting, fawning, etcetera. This disorganized state may underlie increased levels of entropy observed in the overt behavior of severely stressed subjects, which are known to predict the onset of tipping points. Such 'decompensation' or 'dysregulation' can serve as a generic model for episodic mental illness (van de Leemput et al., 2014). In such cases, the hierarchical generative model as shown in Fig. 9B reverts to a hierarchical taxonomy of episodic mental illness ('psychopathology') (HiToP - Kotov et al., 2017). The relative contribution of each phenotypical domain to the overall disease presentation can be parsimoniously expressed as a transdiagnostic factor profile. A differential collapse of the world models of self-functioning, interpersonal functioning and normative functioning should then be a common factor in all forms of mental illness (whether episodic or chronic). The specific type of episodic mental disorder is then determined by the subordinate modules and behavioral primitives that show (disinhibited) disorder as a result of losing these highest levels of integrative control.

This idea is supported by phenotypical studies that show decreased scores on measures of self-functioning and interpersonal functioning as common factors in different forms of mental illness (e.g. Sleep et al., 2019). Also, recent findings show that some changes in brain function are common to a diverse range of acute mental disorders (e.g. major unipolar depression, bipolar disorders, psychosis and anxiety disorders). Such disorders are accompanied by 'transdiagnostic changes' in functional neuroanatomy, which include decreased activity levels in prefrontal and anterior brain regions that support high-level cognitive control (McTeague et al., 2016). These are the same areas that harbor our world models of self, others and global world views (Brunner et al., 2010), which are downregulated under stress (see above). Together, such findings support the idea that all forms of episodic mental illness involve a temporary collapse of higher levels of control that reaches too deeply down the hierarchy. This hypothesis can be tested by studying measures of hierarchical depth in different brain areas as a function of entropy (disorder) levels in hierarchical message passing and corresponding phenotypical changes in healthy controls and patients with different forms of mental illness (see Discussion).

From the above, it follows that individual differences in the degree to which goal hierarchies have grown and matured in the course of life should explain different susceptibilities to mental illness (disorder and tipping points): people with strongly matured hierarchical trees may better withstand the pruning of their hierarchies during a stressful episode than people with lesser developed hierarchical trees. Interestingly, the development of goal hierarchies across the lifespan has been linked to personality development (Russell Cropanzano and Citera, 1993). This process involves the outgrowth and sculpting of goal hierarchies as a result of different forms of associative learning of organisms in relation to themselves and their environments ('maturation'), see above. Whereas episodic mental disorders involve a temporary collapse of goal hierarchies, personality deficits may involve a failure of such structures to develop normally. Neuroimaging studies show that people with (borderline) personality disorders, who are more susceptible to mental decompensation ('crises'), have low volumes of gray matter in the same areas of high-level (cognitive) control that are downregulated under stress (Brunner et al., 2010; McTeague et al., 2016). These underdeveloped brain areas involve the same areas that harbor our world models of self, others and global world views (Brunner et al., 2010). The global faculty of cognitive control that is down-regulated in acute mental illness can therefore be subdivided into high-level world models that support self-functioning (agency), interpersonal functioning (communion) and normative models (meaning), which may each be downregulated to different degrees under stress (Fig. 9B). These mental faculties therefore qualify as 'transdiagnostic factors', which are to some degree involved in all personality disorders (when underdeveloped) and episodic mental disorders (when downregulated).

The collapse of these transdiagnostic world models under stress may cause people to experience specific sets of 'symptoms', i.e. a decreased sense of purpose and normativity (due to collapsing normative functions), a loss of empathic interest in others or the external world, a decreased feeling of communion, derealization (due to a collapse of social / external models), or rather undecisiveness, low self esteem, distorted body-image and/or symptoms such as depersonalization and dissociation (a clinical state characterized by a loss of self-awareness caused by a collapse of self-models). These are typical symptoms of (borderline) patients during acute episodes and may to some degree be common to all patients with mental illness. Indeed, the 'alternative model' for personality disorders in the Diagnostic and Statistical Manual of Mental Disorders (DSM-5) currently lists self-referential and interpersonal functions as two global transdiagnostic factors that are underdeveloped in personality disorders (Zimmermann et al., 2015). These may at some point be supplemented with the third overarching factor (normative functions) as identified in the current paper, a conclusion that is consistent with some existing models of personality development (e.g. Cloninger, 2008). The maturation of these 'great three' world models involves a life-long process of goal-directed learning. The development of these mental domains across the lifespan has been referred to as personality development (or more specifically 'character' formation) (Cloninger, 2008). The various generative models that are subordinate to these three top-level domains (i.e. emotional, motivational, cognitive and executive domains and subdomains) qualify as lesser-order transdiagnostic factors (Fig. 9A). These involve shorter stimulus–response loops that have been associated with Pavlovian learning and habit learning. Such functions develop at earlier stages of life and their stable expression across years has been referred to as 'temperament' (Cloninger, 2008). Individual differences in the expression of such factors are known to produce different personality profiles and susceptibilities for episodic disorders. Together, such findings support the idea that a stress-induced collapse of already underdeveloped regulatory hierarchies triggers disorder and tipping points in human subjects with personality disorders, with shallower hierarchies increasing the risk of such episodes. Future studies may link the hierarchical depth of regulatory hierarchies to scores on specific personality domains and susceptibilities to episodic disorders.

Apart from explaining individual differences in behavior and (susceptibility to) mental illness, the current model may explain individual differences in social interactions. This is because individual organisms can be modeled as hierarchical Bayesian control systems that respond to each other, i.e. the output of one individual (behavior) may serve as the input to another (Friston and Frith, 2015). Such models allow for studies on interpersonal dynamics at small timescales (e.g. stress-induced changes) or at larger timescales (e.g. developmental differences). For instance, a top-down collapse of higher order control may increase extrinsic (social) coupling of one individual with respect to another. This may then cause a collapse of higher order control in the other person (e. g. through a lack of sleep), producing highly recursive (clonic) stimulus–response relationships between two individuals. As a result, two undercontrolled (stressed) individuals may become strongly coupled. This would be a model of strong mutual dependency and/or intense social conflict, including a mutual loss of law-abiding and moral behavior. Much like clonic spinal reflexes that can be silenced only by an external influence, vicious cycles in social behavior are a symptom of insufficient higher-level control that typically require an external party (e.g. mediation, judicial arbitry, or medical intervention) in order to be reduced (Fehr and Fischbacher, 2004).

Thus, individual differences in hierarchical Bayesian control (e.g. personality development) produce stable differences in social interaction, which translate into stereotypical connectivity patterns at a local level, e.g. scores on personality dimensions predict the topological position of individuals in social networks (e.g. Krause et al., 2010). Such individual differences in local connectivity act as simple rules that knit together complex social networks at a global level. This includes the formation of social network clusters in which some opinions and beliefs are held and contrasted with those of other individuals or groups, while trying to get a mutual grip on reality. Social networks may therefore follow similar rules for network architecture and function (collective inference) as shown in Fig. 6.

## 5. Discussion

In the current paper, we present a universal theory on information processing in living systems as well as a general theory on stress and the stress response that are based on first principles in biophysics. We propose that all living systems can be modeled as scale free, *small world* (nested modular) network structures with an information bottleneck structure, resulting in hierarchically organized input (perception), throughput (goal setting) and output (action) parts that are engaged in Bayesian inference. To our knowledge, this is the first time that concepts from network science and graph theory are put together with current ideas on predictive coding to explain hierarchical Bayesian inference in living systems (e.g. Friston et al. (2017)). When embedded in an (a)biotic environment and allowed some freedom of movement, such systems function as 'hierarchical Bayesian control systems', which change their environments through action in order to reduce the difference (error) between their perception of their current inner or outer state (posteriors) and their self-inferred goal states (priors: predictive models of the world of varying levels of complexity). The minimization of this error through either action or model adjustment (learning) corresponds to a gradient search on mean variational free energy, which is known as 'active inference'. Error can be minimized with respect to a hierarchy of goals and corresponding subgoals, with the top of the goal hierarchy representing the most integrated ('highest') goals of the organism. Such goal hierarchies allow organisms to perform an iterative search for ecological niches of a certain predilection, i.e. niche exploration. The prediction error (free energy levels) of hierarchical Bayesian control systems can be defined as 'stress' and the action that follows the error as the 'stress response'. Under stressful conditions, error accumulates vertically in the goal hierarchy and increases the oscillation frequency of network nodes until energy demand exceeds energy supply ('allostatic overload'). The most connected (highest degree, or central) nodes at the top of the information bottleneck (goal hierarchy) are most vulnerable to such energy depletion, causing them to selectively overload and fail. To our knowledge, this is the first explicit mechanistic model of allostatic overload. The selective loss of central (hub) nodes results in a top-down collapse of goal hierarchies, causing organisms to abandon hierarchically higher (more integrated and abstract) goals in favor of hierarchically lower (less integrated and more concrete) goals to save energy and resources. This corresponds to a shift in behavior from long-term, cooperative (prosocial) and/or selfless (altruistic) policies towards short-term, solitary (asocial) and/or self-centered (antisocial) policies. In humans, the (hierarchically) highest goals correspond to social norms and moral values that individuals deem applicable across living systems and timescales. The collapse of such goal states and corresponding behavioral changes under stress corresponds to a blunting of social interactions and, eventually, moral decay (of course, individual differences may cause subjects to deviate from this general rule). Studies indicate that high levels of stress are accompanied by an increase in permutation entropy in measures of hierarchical message passing and overt behavior. Permutation entropy is a measure of 'disorder' in timeseries that quantifies a number of erratic changes. In many living systems, increases in permutation entropy successfully predict a sudden phase transition (a tipping point), indicating disease, or the death of an organism. We propose that an increase in permutation entropy signifies a loss of higher level integrative control across lower-order systems, causing these systems to behave in an uncoordinated, desynchronized (erratic, disordered) way. In humans, a temporary collapse of high-level (integrative) goal states may underlie episodic mental disorders (e.g. major depression, psychosis, panic attacks), whereas a failure of goal hierarchies to mature in the course of life may serve as a model of personality (trait) disorders. The term 'disorder' therefore be seems well-chosen, since it points out an increase in permutation entropy. Whereas a shallow collapse may only cause small changes in behavior, a deeper collapse of goal hierarchies diminishes model complexity to the point where stimuli almost immediately trigger (stress)responses and circularly causal (clonic) patterns emerge between the organism and its environment. Such vicious cycles or attractor states signal a loss of homeostasis and usually align with disease, or the death of an organism. Such changes are universal features of living systems and can be observed at any scale level of organization, including social levels. In order to test these predictions, researchers may need to consider the whole of hierarchical message passing in organisms instead of just parts of it. This has been a major obstacle in the past, but modern data analysis techniques increasingly allow studies of the full complexity of interactions between genes, proteins, metabolites, neurons, brain areas, phenotypes, animal populations and people (the -omics literature). Below, we will discuss several ways of testing these predictions.

### 5.1. General architecture

Our first prediction follows from the universal presence of *small world* topologies in living systems (see Introduction). As a result of this universality, we expect living network systems of any type to show commonalities in network structure. Network structure can be analyzed using software packages such as the igraph library in R (Csardi and Nepusz, 2005) or Cytoscape (Shannon et al., 2003). Small-worldness can be quantified by calculating a small-worldness index, which compares the clustering coefficient (modularity) and average path length of given network to a randomly connected network of equal size (Humphries and Gurney, 2008). A value significantly greater than 1 (and preferably more) indicates that the network is non-randomly connected and contains hub nodes and clusters that allow energy to dissipate along short and efficient paths. Hub nodes can be identified by examining the degree (number of connections) per node, and centrality measures can be calculated that examine the relative importance of nodes in guiding traffic across a network. Hub structures contract their neighboring nodes into network clusters, which can be detected quantitatively by means of network community detection algorithms (e.g. Newman, 2004). Software has been developed that allows detection of so called 'rich club' structures (Fig. 4), which are collections of hub nodes that connect significantly more to other hubs than chance levels (Opsahl et al., 2008). Rich clubs are nested hierarchies of hub nodes that produce a scale invariant network structure. In such structures, each network cluster can be modeled as a node at a next level of spatial aggregation. Functional integration within nested rich clubs structures is an important ingredient of hierarchical Bayesian inference. Also, software packages exist that can test network structures for a diverse range of motifs, e.g. (Masoudi-Nejad et al., 2012). These include bow-tie motifs as well as their constituent motifs, such as feedforward and feedback loops. At the organism level, we expect biological networks to show a nested bow-tie (bottleneck) structure, with cross-connections between similar levels of input and out hierarchies of a (folded) bow-tie, producing input-throughput-output loops of different path lengths. We expect bow-tie motifs to consist of a family of smaller motifs that include feedforward and feedback loops. Studies have already shown an abundance of the feedforward loop motif, which we expect to reflect top-down predictive processing in input hierarchies and bottom-up predictive processing in output hierarchies (Figure 6, Box 1). Such motifs should be counterbalanced by feedback motifs that reflect bottom-up correction of higher-level predictions by lower level prediction errors in input hierarchies (and vice versa in output hierarchies).

With respect to energy flows across biological network structures (network 'function') and its directionality, a distinction can be made between global (macrolevel) and local (microlevel) flows. The input hierarchies of nested bow-tie structures should show multiple excitatory energy streams converging onto higher level hub structures while ascending in the hierarchy, reflecting the functional integration of prediction error signals. Also, input hierarchies should show multiple energy streams diverging while descending in such hierarchy, reflecting top-down and inhibitory predictive control. Together, both information streams reflect perceptive inference. We propose that the directionalities of prediction and prediction error streams are reversed in output hierarchies when compared to input hierarchies, reflecting hierarchical action control (Fig. 6). With respect to local flows, we expect input areas of bow-tie motifs to contain a large proportion of hub nodes with multiple arrows converging onto each hub. Such 'integrator hubs' (sinks, or driver hubs; Yan and He, 2011) are said to have a high in-degree, referring to the number of incoming connections from other nodes that indicate the process of functional integration. Conversely, the output areas of bow-tie motifs should contain a significant proportion of network motifs that involve multiple outputs diverging from a single (hub) node onto a distributed set of other nodes. Such 'distributor hubs' (sources, or driver hubs; Yan and He, 2011) have a large out-degree, referring to the number of outgoing connections that support the process of action control. The throughput parts (knots of bow-ties) may show a substantial number of sources, sinks, and hubs with balanced numbers of incoming and outgoing connections, reflecting continuous cross-evaluation. The net in- and out-degrees of prediction error or predictive hubs are expected to shift along a gradient from input to throughput and output parts of the network, reflecting a smooth transition between these domains. As observed, we expect the dynamics of bottom-up and top-down units (as well as within-level dynamics) to produce oscillatory behavior of different spatiotemporal scale, i.e. attractor states.

Predictions with respect to the directionality of links in biological networks can be tested using software developed to study causal relationships (conditional dependencies in time) between mutually dependent variables (e.g. Scheines et al., 1998). Inferring directions amongst variables using causal reasoning software is considered a hard problem in statistics and the directions obtained may not always be reliable. In networks in which nodes have clear and measurable relationships (e.g. genomic, proteomic or neural networks), it may be quite feasible to infer directions, whereas in other networks (e.g. statistical networks used to study brain function or phenotypical states), testing these hypotheses may prove to be more difficult. Recent attempts to infer both global (Hillebrand et al., 2016) and local (Märtens et al., 2017) directionalities of functional connections in the human brain have involved the use of a novel and promising phase transfer entropy measure (Lobier et al., 2014). Interestingly, this measure may partly reflect the flow of prediction error (free energy) across network systems, since entropy and energy measures are related through the second law of thermodynamics. Using phase transfer entropy, a bottom-up convergence was demonstrated in sensory areas, consistent with ascending prediction errors within the first part of a bow-tie structure. Evidence for top-down divergence was less clear, however. At a more local level, a bidirectional convergence/divergence motif was found, which may reflect true bidirectionality or an insufficient decomposition of flow directionalities into ascending (excitatory) prediction errors and descending (inhibitory) predictions. Overall, the quantification of energy flows and their directions within biological networks is an important venue for further study. Similar measures that are used to study directionality of energy flows in brain function can be used to study molecular or neural networks.

Several of the predictions made in this paper require a quantification of the concept of 'hierarchy'. Despite its common use in everyday language, it has proven a challenge to produce a formal definition of hierarchy, hence several definitions exist (Corominas-Mutra et al., 2013). In *small world* networks, some nodes or clusters may only exist by virtue of other nodes or clusters, i.e. they form conditional dependencies in space (a hierarchy of part–whole relationships; Ravasz and Barabasi, 2003). Additionally, biological networks involve state changes that follow a hierarchy of conditional dependencies in time (i.e. causal order, or directionality). Both hierarchies need to be accounted for in order to obtain an idea of the hierarchical order of nodes or clusters in scale invariant network structures. Perhaps the most formal definition of hierarchy is provided by Corominas-Mutra et al. (2013). The authors propose to quantify hierarchy in terms of three key elements, which include treeness (pyramidical shape, or spatial hierarchy), feedforwardness (top-down or bottom-up directionalities, or temporal hierarchy) and orderability (the effect of causal cycles), allowing the hierarchical structure of different types of networks to be directly compared within a single three-dimensional space. This definition of hierarchy controls for the nestedness and directionality of links, but needs to be adapted for weighted networks. Perhaps a more straight-forward approach to measuring the number of hierarchical levels of a biological network structure would be to count the number of nested relationships between clusters and subclusters (i.e. scale levels) regardless of directionality (Kaiser and Hilgetag, 2010). The number of functionally segregated subclusters that are integrated in a nested fashion into a particular hierarchy of control provides a measure of the height of a hierarchical tree (Newman and Girvan, 2004). Several hierarchical network clustering algorithms exist that can provide information on the number of part- whole relationships, allowing for the construction of corresponding tree-graphs (e.g. Lancichinetti and Fortunato, 2009). Measures of nestedness (hierarchical depth) should be intimately tied to the proportion of functional integration versus segregation of network clusters. This relationship can be tested quantitatively by using another measure derived from neuroscience, called neural complexity (CN; Rubinov and Sporns, 2010; Tononi et al., 1994). This measure defines functional segregation as the relative statistical independence of small clusters of a system and functional integration as significant deviations from independence of larger clusters. CN expresses the average deviation from statistical independence for clusters of increasing size. CN values are high when functional segregation and integration coexist in a balanced manner and low when the components of a system are either completely independent (segregated) or completely dependent (integrated). Although first used to analyze neural networks, this measure captures a universal feature of biological systems (Rubinov and Sporns, 2010). Although CN is a structural measure, it may well serve as a means to quantify Bayesian model complexity, which involves the number of independent variables (degrees of freedom) that are available to a particular model. Model complexity is expected to decrease when moving up the hierarchy of generative models, since higher level models offer a more parsimonious explanation of lower-order events (Spiegelhalter et al., 2002). Other measures to quantify information integration and corresponding fitness have been suggested as well, e.g. Edlund et al. (2011). Together, these measures of (nested) hierarchical depth and model complexity can be used to test predictions with respect to the comprehensiveness of hierarchical control in biological networks (see previous sections for such predictions). Briefly, we expect the amount of functional integration across multiple contextual cues (and the corresponding height of the nested hierarchical tree) to differ between lower (less) and higher (more) organisms, and individuals or species with lower (less) or higher (more) levels of autonomy/agency and self-directedness, solitary (less) and more social (more) behavior, less (less) and more (more) *prosocial* behavior, smaller (less) and larger (more) amounts of parental investment, less (less) or more (more) transgenerational awareness and actions, less (less) and more (more) normative (law abiding) or moral behavior, and between calm (more) and stressful (less) situations (see below). Such differences may involve specific parts of the network, e.g. throughput hierarchies may show greater (individual) differences in hierarchical depth than perceptive or output hierarchies.

As discussed, hierarchical depth is related to the ability of an organism to control its internal states or the world around it. Organisms with lesser developed hierarchies may therefore find it more difficult to adapt to complex and changing environments. In the specifically human case, the maturation of deep goal hierarchies in humans can be linked to personality development, and insufficient maturation of hierarchical trees to personality disorders and instability (mental illness). Such deficits eventually decrease scores on measures of self models (agency), social models (communion) and normative models (meaning). Future studies may compare the hierarchical network structure of subjects with and without personality disorders to further test these predictions, e.g. using neuroimaging techniques. As observed in Section 4, individual differences in the height or maturation of goal hierarchies can also be linked to stable individual differences in social interaction, which define the local topology in social networks to eventually affect the global structure of social networks.

As a general remark, hierarchical Bayesian inference describes a mechanism for inferring 'signs out of signs', which amounts to a model of semiotics (Fortier and Friedman, 2018). Social connections can be defined in terms of the exchange of free energy between different agents through synchronized action-perception cycles and have produced a novel way of thinking about reciprocity and hermeneutics (Friston and Frith, 2015; Vasil et al., 2020). Organisms may act in such a way as to alter the amount of free energy (model error, stress) in other beings. This corresponds to aiding other organism with information or hampering them by not sharing information or providing desinformation, which has a strong moral connotation. Indeed, our model predicts that organisms and people that produce the most detailed and accurate models of the world are at a thermodynamic disadvantage when operating alone (but at a significant advantage when working together). The current paper sees hierarchical Bayesian inference as a way to explain our highest levels of mental functioning, including the formation of social norms and moral goals. Individuals may differ in the degree to which such models have developed and therefore differ in the degree to which their behavior is guided by higher moral principles. Such topics have been kept to the realms of philosophy for many thousands of years. Especially as regards moral functioning, one should be careful not to commit to a naturalistic fallacy by assuming that the factual structure and dynamics of biological systems automatically informs us of a desirable structure (Moore and Baldwin, 1993). Although one should be prudent, however, it is not impossible to move from facts ('is') to moral prescriptions ('ought'), especially when such facts involve things of a hierarchical generative and symbolic nature (i.e. humans as symbolic animals). The relative autonomy of high-level generative models with respect to the lower-level events from which they have been inferred makes it possible to produce highly creative models that go well beyond the available facts (predictions, goals), yet still have their basis in such facts (memories). This relative disconnection may be what is required to finally integrate science and morality safely within a single discipline (the 'moral sciences;' Ruse, 1988). This being said, it may well be a 'categorical imperative' for all people to actively develop mature regulatory hierarchies that incorporate as many contextual cues as possible into self-transcending world models that allow our behavior to be informed by universal laws and social standards through which people may connect (predict each other) across nations, cultures and timescales. Our scientific, legal and moral institutions may facilitate the exchange of such commonly held values or goals in order to facilitate belief updating and develop a commonly held world view. The detrimental effects of (chronic) stress on such a development should be actively countered across many generations.

## 5.2. Stressful conditions

Organisms continuously change their wiring patterns while anticipating and responding to different situations. This produces a dynamic balance between the functional segregation and integration of network communities and, therefore, hierarchical structure (Sporns, 2013). In this paper, we propose that severe stress alters network community structure of biological systems in a universal way, i.e. it should produce a shutdown of high-level hub structures within the information bottlenecks of organisms (i.e. the knots of bow ties). This shifts the balance between functional integration and segregation towards functional segregation, thereby reducing the height of the nested hierarchical tree (see above). Such changes may not cause a significant shift in the small-worldness measure, but may decrease hierarchical depth as measured by hierarchical clustering algorithms. Thus, severe stress should produce shorter and shallower bow-tie motifs with wider knots, which interferes with the ability of organisms to compress information. This should translate into increasingly shorter loops that run from input via processing to output parts of a (functional connectivity) network. This can be tested by measuring the path length measure that run from input to output structures for different nodes of interest (i.e. the average distance from one node to another via a subset of intermediate nodes). We expect measures of hierarchical depth to be high in moderately stressful situations and low under either very low or very high levels of stress (i.e. either complete segregation or integration). Additionally, we expect stress to shift the balance between predictive processing and corrections of such predictions in favor of prediction errors, making organisms more susceptible to belief updating by immediate sensory evidence. Such changes involve an increase in the synchronous gain (precision) of prediction error signals versus predictive signals, which involve changes in connective efficacy e.g. as a result of (neuro)modulatory signaling pathways in neural or molecular networks. The overall result of such changes may be examined by measuring shifts in scores on measures of directed and weighted connectivity, e.g. phase transfer entropy studies showing increase bottom-up convergence as opposed to top down divergence in perceptive or goal hierarchies and vice versa in output hierarchies.

Finally, when stress levels are particularly high, we expect tell-tale signs of undercontrolled control systems in the form of increased permutation entropy (or critical slowing down) and changes in overt behavior that signify a reduction in model complexity. This can be tested by linking entropy levels and tipping point thresholds to measures of hierarchical depth and behavioral changes in different individuals or species. Such studies are readily performed in bacteria and other microbes, where e.g. acidity, salinity or antibiotic levels may be varied to examine bacterial responses in hierarchical message passing and growth or survival rates (Nagar et al., 2016; Marles-Wright et al., 2008; Yu and Gerstein, 2006; Zhu et al., 2020). For obvious reasons, however, such studies cannot be easily translated to higher organisms. Actively bringing sentient creatures to the brink of a tipping point would be highly unethical. In the specifically human case, severe stress does appear to decrease the amount of functional integration within the human brain, as measured by an information processing efficiency measure (Rubinov and Sporns, 2010; Wheelock et al., 2018). Another study in post-traumatic stress syndrome reports increased amounts of functional segregation (Zhu et al., 2019). Yet other studies show that the human brain falls back from goal- directed to habitual control during stress (Schwabe and Wolf, 2009, 2011). Such findings are in line with a collapse of high-level integrative control, but require a systematic approach in order to prove the principles put forward in this paper. Although experimental studies are precluded, however, studies of mental illness may provide a natural situation in which to examine tipping point thresholds in relation to hierarchical depth in humans. As observed, we expect individual differences in the hierarchical depth of goal hierarchies to explain individual differences in resilience and susceptibility to mental disease. Such hierarchies ultimately involve highly integrated self models, social models and transcendent world models. A temporary collapse of these models should be a common (transdiagnostic) factor in all episodic forms of mental illness ('psychopathology)'. Conversely, a persistent failure of similar cortical hierarchies to mature properly should underlie a stagnation of personality development and the concomitant chronic risk of episodic mental illness ('personality pathology'). In terms of diagnostics, monitoring scores of patients on these three global domains therefore seems crucial. In terms of prognostics, measuring change scores on these three global domains may help to predict treatment success and relapse rates, whereas entropy levels in ESM timeseries may help to predict the onset of episodic mental disorders (i.e. tipping points, van de Leemput et al., 2014). In terms of therapeutics, promoting an optimal balance between top-down predictions and bottom-up belief updating can be performed by prescribing medication that modulates synaptic gain (e.g. antidepressants, antipsychotics, etcetera) or by means of transdiagnostic psychotherapeutic interventions that promote the updating of false beliefs with respect to self, others, and global world views (e.g. exposure and re-appraisal in cognitive behavioral therapy). Of course, much can be won by prevention strategies that discourage people from developing maladaptive world models in the first place, e.g. by providing children with a safe social environment, proper training and education.

To summarize, we expect stress to alter (functional) connectivity in living systems in canonical ways, regardless of whether that involves single-cellular life forms of complex multicellular organisms and higher species. This allows for the categorization of stress-levels into discrete stages, each with distinct and quantifiable features (for a similar attempt, see Romero et al., 2009). 'Low' amounts of stress (prediction error) should be associated with low-level action–perception cycles, i.e. activity of short loops within the nested hierarchy and low levels of permutation entropy in hierarchical message passing ([0] Reflexive, habitual behavior, homeostatic control). This reflects the successful suppression of low-level prediction errors by predictive structures of low model complexity (short stimulus-evaluation-response loops). When stress levels rise to mild or moderate levels, we expect increased involvement of higher level generative models and behavioral policies of corresponding complexity that conspire to suppress rising levels of prediction error. This stage involves increased activity within processing loops of increasing length and rising levels of permutation entropy ([1] goal-directed behavior, allostatic control). In contrast, we expect the activity of higher hierarchical levels to decrease again when stress levels become more severe. This reflects the dissolution of higher-level goal states when the hierarchy is taxed to its limits as a result of hub overload and failure ([2] regression to homeostatic behavior, allostatic overload). Thus, both low [0] and high [2] stress levels should engage habitual rather than goal-directed forms of behavior. The final two stages involve an undercontrolled state of high permutation entropy (([3] critical slowing down, CSD), which predicts a sudden loss of functional or structural integrity ([4] loss of control, tipping points / decompensation). All of these stages can be identified objectively (van de Leemput et al., 2014; Zhu et al., 2020). Predictions with respect to (the directionality of) network connections and levels of disorder in hierarchical message passing under different levels of stress can be tested using the quantitative measures described above.

An interesting approach would be to simulate changes in the performance of hierarchical control systems under different levels of stress using artificial systems (e.g. information bottleneck systems, see below). Such studies would allow testing the hypothesis that folded bow-tie structures with nodes that collapse as a function of node degree as a result of rising levels of prediction error would show a top-down collapse of goal hierarchies (the knots of bow ties), i.e. to produce a model of allostatic overload. This can be done by modeling precision as the weight of connections of prediction error units relative to predictive units. Increases in entropy measures of both hierarchical message passing and behavioral sequences produced by the system should then be predictive of tipping points / loss of homeostasis and serve as a universal model for stagnant growth, disease, or death. Such studies would be a safe way to study the tipping point thresholds as a function of hierarchical depth, providing a generic model for individual differences in fitness. Incidentally, such a model would provide a mechanistic account of the workings of natural selection on organisms that lack adaptive capacity and, thus, link to studies of evolutionary biology.

Finally, it would be interesting to examine to what degree the structure of human phenotypical networks (inner experience and overt behavior) echoes the physical structure of living network systems as shown in Figs. 6 and 8. Phenotypical networks indeed show signs of small-worldness and nested modular hierarchy (part–whole relationships), as well as statistical dependencies between items that can be explained by physical network architectures capable of hierarchical Bayesian inference (e.g. Goekoop and Goekoop, 2014). A similar approach can be tried in social networks. Here, agent-based simulations could aid in understanding patterns of social interaction at the local level (e.g. mutual dependence or social conflict) as well as global phenomena such as innovation and rumor diffusion, voting, migration, strikes, riot behavior, economic slowdown and warfare.

### 5.3. Modeling organisms: a unified theoretical framework

One of the most interesting features of living systems is that they follow scale-independent rules of network structure and function that apply to all organisms. Such universality means that organisms of any type can be modeled using a minimum set of building blocks under a common theoretical framework. Scholars will not have to make unique models for each organism separately, nor for each level of observation within the organism (e.g. genetic, cellular, systems level, or social). Instead, organisms can be described in terms of a limited set of network motifs (Araujo and Liotta, 2018), allowing for compression of datasets. Additionally, scale invariance means that organisms can be modeled either in all their intricate detail (i.e. the full hierarchy of part-whole relationships) or rather more grossly, as a few global motifs that together perform some global functions, without losing too much information (Fig. 10). Such multilevel 'coarse graining' techniques have been shown to be successful in simulating organismic behavior (Derbal, 2013). Here, we show that organisms can be grossly modeled as a giant predictive feed forward loop (FFL; section 3.3), which produces output that provides an update on these predictions via the environment (active inference).

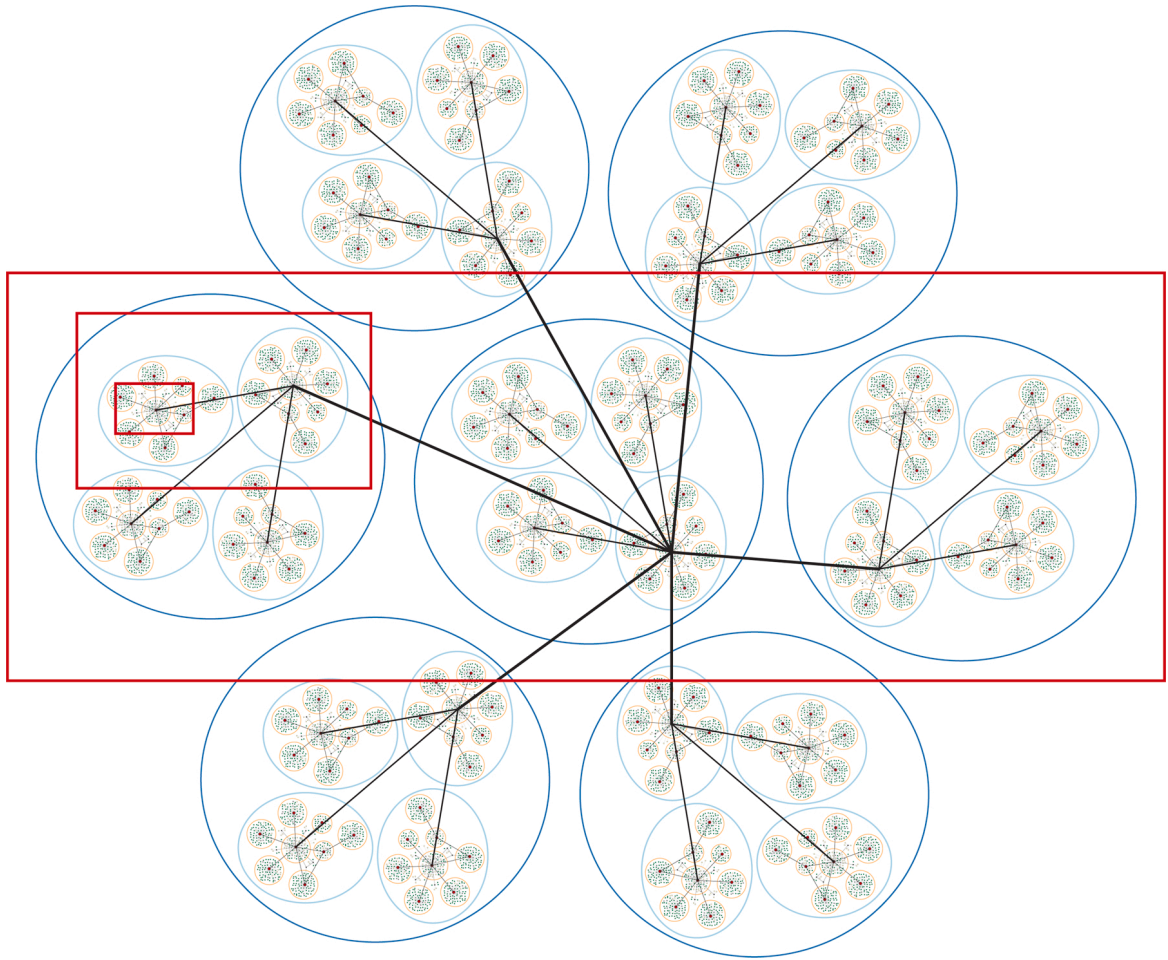

**Fig. 10.** Scale-Invariant Features in Organisms Allow for Efficient Modeling.
*Note:* The scale invariance of biological networks proves useful for modeling organisms. The same network motifs appear at different scale levels of organization, where they support similar functions. For example, red squares indicate the same structural network motif (a folded bow-tie motif) at different spatial scale levels of observation. Modeling organisms would only require knowledge of the number, positions and scale levels of a particular type of motif within an organism, allowing for significant parsimony of description (i.e. organisms can be 'compressed' and 'decompressed'). The fractal-like structure of biological networks means that organisms can be reduced to single feedforward loops at the highest spatial scale level of observation (the level of the individual organism) without losing much information. See text for further details.

### 5.4. How biology may inform machine learning

So far, we have discussed how artificial intelligence can help us to understand biological networks in terms of hierarchical Bayesian control systems. Conversely, one may examine how biological systems may inform computer models of hierarchical control systems. For instance, deep networks usually start out with random connections that change after learning. Eventually, the idea of learning is to connect some input (e.g. a series of pixels that together form the shape of a cat) to a desired output (say, the succession of letters (C-A-T) in a non-random fashion by means of a hierarchically organized throughput area that makes these connections. We have seen that such associations are significantly improved when allowing for a hierarchical structure of input, throughput and output modules (Section 2.2). Since non-randomly wired *small world* networks form spontaneously when optimizing the flow of energy through random networks (Jarman et al., 2017), we predict that present-day hierarchical deep networks, when performing at optimal efficiency, must have approached a scale invariant, *small world* network structure. Currently, we know of no studies that have examined existing deep networks directly for small-worldness. A recent study found that fitting a deep network with *small world* network architecture prior to learning significantly enhanced its performance, thanks to the rapid convergence of microstates onto hub states (Javaheripi et al., 2019). A further improvement could be made by fitting deep networks with bottleneck (bow-tie) structure prior to learning (Shwartz-Ziv and Tishby, 2017). Several studies show that information bottlenecks increase the performance of hierarchical (deep) networks by allowing their higher hierarchical levels to perform some kind of compression and generalization of events that take place at lower levels (Hafez-Kolahi and Kasaei, 2019; Shwartz-Ziv and Tishby, 2017). Such performance increases appear to be related to phylogenetic learning (evolution) rather than ontogenetic learning (within-lifespan individual development), hence their introduction may significantly boost system performance by skipping a generic (phylogenetic) learning process, allowing the system to directly proceed with task-relevant (ontogenetic) learning instead. Information bottlenecks may also prove crucial in studies of hierarchical Bayesian inference (interestingly, the objective function used for the free energy principle, i.e. variational free energy, can be cast in terms of compressing and minimum description lengths (Friston, 2019a; MacKay, 1995, 2003; Sun et al., 2011; Wallace and Dowe, 1999). Given the ubiquitous presence of *small world* and bottleneck networks in nature, we expect that such features will soon be detected in hierarchical deep learning systems and that the formation of such structures correlates positively with the performance of such systems. Indeed, the very structure of deep networks necessarily entails a kind of bowtie structure. This is most evident in things like variational autoencoders, which arguably represent the state-of-the-art in deep learning (Zhao et al., 2017). These are deep networks with a bow-tie like architecture that follow the rules of hierarchical Bayesian inference, with a converging input part that is called an 'encoder' and a divergent output part that is called a 'decoder'. Behavior is generated by decoding abstract states into hierarchical output sequences in a top-down manner. We predict that such structures will show biologically plausible behavior when folded to connect input and output structures at corresponding hierarchical levels (Safron, 2020) and when accounting for hub overload and failure during stress (Stam, 2014). Overall, it is interesting to note that the network architectures that predominate in machine learning (e. g. deep convolution neural networks) conform almost exactly to the principles that we have been exposing, i.e. they have an explicit hierarchical structure with a certain kind of sparsity, following rules of predictive coding and hierarchical Bayesian inference.

As a final remark, biological systems may inspire machine learning techniques with respect to the generic response they show to severe stress and the overtaxing of their hierarchies of control. Lowering integrative control at the cost of contextual integration may be an answer in situations that require rapid decisions within the context of limited energy supply (e.g. battery powered devices). This may speed up system performance in dire situations, e.g. when used in military situations, self driving cars or policing. The prospect of 'stressed robots' that weigh selfish and selfless goals may not seem very appealing, but may ultimately prove to be of significant value. For instance, robots may be programmed to never abandon higher level (normative) goals over lower level (self-centered or social) goals in relevant situations, effectively causing them to remain morally just and impartial, or to self-sacrifice (fail for the global good) under stressful conditions.

### 5.5. Conclusion

To conclude, we have examined how biological network systems have structural features that allow them to function as hierarchical Bayesian control systems. Such systems have generic ways of producing behavior and responding to stress, which may prove useful in understanding animal as well as human behavior. Biology on the other hand keeps on inspiring man-made systems, for which we have made some suggestions. A list of techniques has been presented that can be used to test the hypotheses presented in this paper.

## Acknowledgements

We thank the reviewers for their constructive comments that added significantly to the quality of the paper.